\documentclass[useAMS,usenatbib]{mn2e}

\usepackage{epsfig}
\usepackage{amssymb}
\usepackage{natbib,multirow}

\def\deg{\ifmmode^\circ\else$^\circ$\fi}
\def\Msun{\ifmmode{\mathrm M_\odot}\else{M$_\odot$}\fi}

\def\ah{\ifmmode{^\textrm{\scriptsize h}}\else{$^\textrm{\scriptsize h}$}\fi}
\def\am{\ifmmode{^\textrm{\scriptsize m}}\else{$^\textrm{\scriptsize m}$}\fi}
\def\as{\ifmmode{^\textrm{\scriptsize s}}\else{$^\textrm{\scriptsize s}$}\fi}

\title[Evolutionary paths among different red galaxy types at $0.3<z<1.5$]{Evolutionary paths among different red galaxy types at $0.3<z<1.5$ and the late buildup of massive E-S0's through major mergers}

\author[Prieto et al.]
{\parbox{\textwidth}{Mercedes Prieto$^{1,2}$\thanks{Affiliations are listed at the end of the paper.}\thanks{E-mail:
mpm@iac.es}, M.\,Carmen Eliche-Moral$^{3}$, Marc Balcells$^{4,1,2}$, David Crist\'obal-Hornillos$^{5,6}$, Peter Erwin$^{7,8}$, David Abreu$^{1}$, Lilian Dom\'{\i}nguez-Palmero$^ {1}$, Angela Hempel$^{1,2}$, Carlos L\'opez-Sanjuan$^{5}$, Rafael Guzm\'{a}n$^{9}$, Pablo G.\,P\'erez-Gonz\'alez$^{3,10}$, Guillermo Barro$^{3}$, Jes\'{u}s Gallego$^{3}$, and Jaime Zamorano$^{3}$}}

\begin{document}

\date{Accepted 2000 ... ... . Received 2000 ... ...; in original form \today}

\pagerange{\pageref{firstpage}--\pageref{lastpage}} \pubyear{2000}

\maketitle

\label{firstpage}


\begin{abstract}

Some recent observations seem to disagree with hierarchical theories of galaxy formation about the role played by major mergers in the late buildup of massive E-S0's. We re-address this question by analysing the morphology, structural distortion level, and star formation enhancement of a sample of massive galaxies ($\mathrm{M}_*> 5\times 10^{10}\Msun$) lying on the Red Sequence and its surroundings at $0.3<z<1.5$. We have used an initial sample of $\sim$1800 sources with $K_s<20.5$ mag over an area $\sim 155$\,arcmin$^{2}$ on the Groth Strip, combining data from the Rainbow Extragalactic Database and the GOYA Survey. Red galaxy classes that can be directly associated to intermediate stages of major mergers and to their final products have been defined. We report observational evidence of the existence of a dominant evolutionary path among massive red galaxies at $0.6<z<1.5$, consisting in the conversion of irregular disks into irregular spheroids, and of these ones into regular spheroids. This result implies: 1) the massive red regular galaxies at low redshifts derive from the irregular ones populating the Red Sequence and its neighbourhood at earlier epochs up to $z\sim 1.5$; 2) the progenitors of the bulk of present-day massive red regular galaxies have been disks that seem to have migrated to the Red Sequence mostly through major mergers at $0.6<z<1.2$ (these mergers thus starting at $z\sim 1.5$); and 3) the formation of E-S0's that end up with $\mathrm{M}_*> 10^{11}\Msun$ at $z=0$ through gas-rich major mergers has frozen since $z\sim 0.6$. All these facts support that major mergers have played a dominant role in the definitive buildup of present-day E-S0's with $\mathrm{M}_*> 10^{11}\Msun$ at $0.6<z<1.2$, in good agreement with hierarchical scenarios of galaxy formation.
\end{abstract}


\section{Introduction}
\label{Sec:introduction} 

Studies based on data from the Sloan Digital Sky Survey (SDSS) have revealed the existence of a color bimodality in the mass distribution of local galaxies \citep{2001AJ....122.1861S,2003MNRAS.341...33K,2004ApJ...600..681B}. The most massive systems (basically spheroids) accumulate into a well-defined Red Sequence in color-magnitude diagrams, while the less massive blue ones (mostly disks) reside into a more spread Blue Galaxy Cloud  \citep{2001AJ....122.1861S,2003MNRAS.341...54K,2004ApJ...600..681B}. Although this color-mass bimodality is observed up to $z\sim 1$, observational data evidences a strong evolution in it for both field and cluster environments \citep{1998ApJ...497..188C,2000ApJ...541...95V,2001ApJ...553...90V,2004ApJ...608..752B}. In fact, the mass limit isolating the Red Sequence from the Blue Cloud rises as the redshift increases and the stellar mass harbored by the former has nearly doubled since $z\sim 1$ \citep[whereas that of the Blue Cloud has remained nearly constant, see][]{2007A&A...476..137A,2008ApJ...672..177L,2008ApJ...677..828S,2009ApJ...694.1171T}. This points to a progressive buildup of the Red Sequence during the last $\sim 9$\,Gyr, associated with the migration of disks from the Blue Cloud to the Red Sequence through mechanisms that are still poorly understood, but that must be responsible of their star formation quenching and morphological transformation \citep[][F07 hereafter]{2000ApJ...536L..77B,2007ApJ...665..265F}.

According to hierarchical models of galaxy formation, the mechanism governing this evolution in the most massive systems have been major mergers \citep[i.e., with mass ratios from 1:1 to 1:4, see][]{1999MNRAS.310.1087S,2002NewA....7..155S}. Present-day massive spheroids (E-S0's) are expected to be the result of the most massive and violent merging sequences in the Universe, also being the latest systems to be completely in place into the cosmic scenario  \citep[at $z\lesssim 0.5$, see][]{2006MNRAS.366..499D,2008ApJS..175..356H,2010ApJ...725.2312O}. However, this prediction conflicts directly with recent data indicating that massive galaxies seem to have been in place before their less-massive counterparts \citep[a phenomenon known as downsizing, see][]{2006ApJ...651..120B,2006A&A...453L..29C,2008ApJ...675..234P,2011arXiv1101.2867M}. 

Also, this assembly epoch seems to depend strongly on the galaxy mass and its environment \citep[][]{2005ApJ...621..673T,2005A&A...442..125D,2006ApJ...652L.145D,2006ApJ...647L..99D,2008ApJ...687...50P,2010MNRAS.402.1942C,2010MNRAS.405..477N,2010MNRAS.401L..39V}. The wide range of ages found for E-S0's ($\sim 2$-15\,Gyr) and the disagreement between optical and NIR age estimators (sometimes, of up to $\sim 6$\,Gyr for a given galaxy) indicate that E-S0's have been built up at different epochs and through different mechanisms, basically depending on their masses \citep[][]{2000AJ....119.1645T,2006ApJ...647..265B,2010AJ....139..540L}. It is generally accepted that the time period at $1<z<2$ is a transition era in which an increasing fraction of galaxies end their star formation activity and move onto the Red Sequence (\citealt{2007A&A...476..137A}, but see also \citealt{1997ARA&A..35..389E, 2003ApJ...595...71C, 2006ApJ...639..644E,2009ApJ...694.1171T,2011arXiv1104.2595B,2011MNRAS.417..900D}). Nevertheless, there are conflicting views on the number evolution experienced by massive E-S0's at $z\lesssim 1$, with studies reporting from a negligible evolution to an increase by a factor of up to $\sim 3$ in these systems \citep[][]{2006A&A...453..809I,2007ApJ...665..265F,2007ApJS..172..406S,2008ApJ...682..919C,2011ApJ...727...51N}. 

Moreover, the Red Sequence is not made of a homogeneous galaxy population, but of a mixing of different galaxy types that has evolved strongly with redshift \citep[][]{2002A&A...381L..68C,2003ApJ...586..765Y,2003MNRAS.346.1125G,2004ApJ...600L.131M,2007A&A...465..711F,2011MNRAS.414.2246H}. In the last years, different studies have analysed the number evolution of the Red Sequence from color- and morphologically-selected samples, but the conclusions derived from the former do not apply to the latter ones in general \citep{2007A&A...465..711F,2010ApJ...719.1969B,2011MNRAS.414.2246H}. Recently, \citet{2010ApJ...709..644I} have considered simultaneously both the morphological and star-formation properties of red galaxies to analyse their buildup. These authors confirm that the bulk of massive quiescent galaxies is rapidly created at $1\lesssim z\lesssim 2$, this mass assembly becoming negligible at later epochs. As in many previous studies, they also propose major mergers as drivers of this buildup. 

Several estimates assuming that most of the mass growth in quiescent galaxies is due to mergers indicate that this mechanism is capable of explaining at least 50\% of the number density evolution of massive galaxies \citep[see, e.g.,][]{2010A&A...519A..55E,2010arXiv1003.0686E,2011ApJ...739...24B}. However, no direct observational evidence has been found up to the date on the existence of a cause-and-effect link between major mergers and the definitive assembly of massive quiescent galaxies \citep[][]{2007ApJ...665..265F,2009MNRAS.398...75S,2011MNRAS.411..675S,2010MNRAS.tmpL.176B,2011MNRAS.tmp...90B,2010ApJ...718.1158L,2011MNRAS.413.1678C}. 

The present study tries to advance in the understanding of the formation and evolution of massive galaxies by analysing the physical properties of red galaxies at $0.3<z<1.5$ with stellar masses $\mathrm{M}_*>5\times 10^{10}\Msun$. The novelty of our study over previous ones is two-fold. First, we include information about the structural distortion of each galaxy to trace merger remnants (besides considering morphology and star formation properties). And secondly, as most objects in their evolution towards the Red Sequence must have gone over nearby Green Valley locations transitorily (F07), we have analysed the red galaxies both lying on the Red Sequence and at close positions on the Green Valley. Therefore, red galaxies in the context of this paper include both the galaxies on the Red Sequence and at its neighbourhood. The galaxy classes resulting from the combination of morphological, structural, and star-formation activity properties allow us to trace the evolution of intermediate stages of major mergers and of their final remnants since $z\sim 1.5$. Finally, the observed number density evolution experienced by each galaxy type is used to carry out a set of novel observational tests defined on the basis of the expectations of hierarchical models, which provide observationally and for the first time  main evolutionary paths among the different red galaxy types that have occurred in the last $\sim 9$\,Gyr. 

The paper is organized as follows. In \S\ref{Sec:sample}, we provide a brief description of the survey. Section \S\ref{Sec:RGSelection} is devoted to the definition of the mass-limited red galaxy sample. In \S\ref{Sec:classification}, we define the galaxy classes according to the global morphology, structural distortion level, and star formation enhancement of the red galaxies. In \S\ref{Sec:Error}, we comment on the sources of errors and uncertainties. Section \S\ref{Sec:Tests} presents three novel tests to check the existence of any evolutionary links between the different red galaxy types, based on the expectations of hierarchical models of galaxy formation. The results of the study are presented in \S\ref{Sec:Results}. In particular, the results of the three tests proposed for the hierarchical scenario of E-S0 formation can be found in \S\ref{Sec:TestsResults}. The discussion and the main conclusions of the study are finally exposed in \S\S\ref{Sec:Discussion} and \ref{Sec:Conclusions}, respectively. Magnitudes are provided in the Vega system throughout the paper. We assume the concordance cosmology \citep[$\Omega_\mathrm{m} = 0.3$, $\Omega_\Lambda = 0.7$, and $H_0 = 70$\,km s$^{-1}$ Mpc$^{-1}$, see][]{2007ApJS..170..377S}.

\begin{figure}
\includegraphics[width=0.5\textwidth,angle=0]{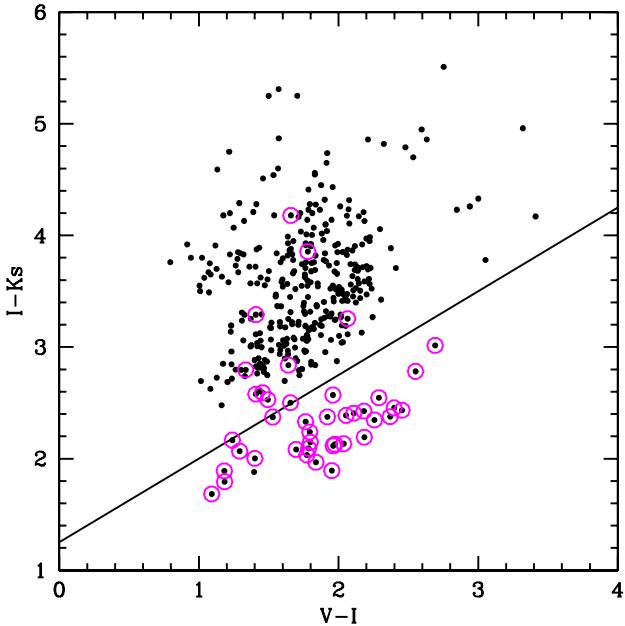}
\caption{Color-color diagram for distinguishing stars from galaxies  ($I-K_{s}$ vs.~$V-I$). \emph{Dots}: Data from the original $K$-band selected catalogue. \emph{Solid line}: Color cut defined to isolate stars (i.e., the sequence of data located below the line) from galaxies (data lying above it). \emph{Circles}: Objects classified as "stars" or "compact" visually (see \S\ref{Sec:Morphology}). All objects identified as "stars" visually are located in the stellar region of this diagram. }\label{Fig:star-galaxy}
\end{figure}

\begin{figure}
\includegraphics[width=0.5\textwidth,angle=0]{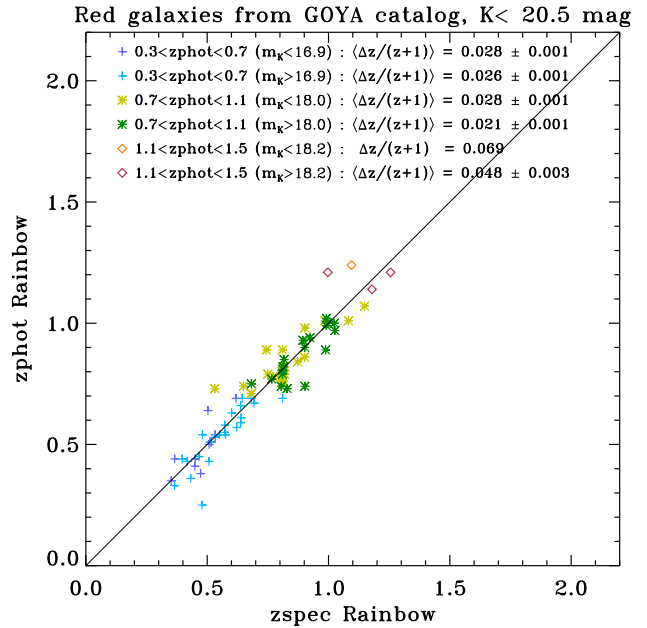}
\caption{Spectroscopic redshifts vs. photometric redshifts for the red galaxies in the sample having spectroscopic confirmation in the DEEP2 catalog \citep{2003SPIE.4834..161D,2007ApJ...660L...1D}. Bright and faint galaxies in each one of the three wide redshift bins under consideration in the study are plotted with different symbols. The typical photometric redshift uncertainty for the red galaxy sample is below $\Delta(z) / (1+z) <0.03$ at all redshifts for both bright and faint sources (see the estimates in the figure).}\label{Fig:zspec-zphot}
\end{figure}

\section{The sample}
\label{Sec:sample}

We have combined multiwavelength data from the Rainbow Extragalactic Database\footnote{Rainbow Extragalactic Database:\\https://rainbowx.fis.ucm.es/\-Rainbow\_Database/\-Home.html} \citep{2011ApJS..193...13B,2011ApJS..193...30B} and the GOYA photometric survey\footnote{GOYA project (Galaxy Origins and Young Assembly):\\ http://www.astro.ufl.edu/GOYA/home.html} \citep[see][]{2002INGN....6...11B} over an area of $\sim 155$\,arcmin$^{2}$ in the Groth Strip \citep[$\alpha=14^h 16^m 38.8^s$ and $\delta=52^o 16' 52''$, see][]{1994AAS...185.5309G,1999AJ....118...86R,2002ApJS..142....1S}. The Rainbow Extragalactic Database compiles multi-wavelength photometric data from the UV to the FIR (and, in particular, in Spitzer/MIPS 24$\mu m$ band) over this sky area, providing analysis of spectral energy distributions of nearly 80,000 IRAC 3.6$+$4.5 $\mu m$ selected galaxies. This study considers the photometric redshifts available in the Rainbow Database, which have a typical photometric redshift accuracy of $<\Delta z/(1+z)> = 0.03$ \citep{2011ApJS..193...30B}, as derived for the sources with spectroscopic redshifts available in the DEEP2 Galaxy Redshift Survey \citep{2003SPIE.4834..161D,2007ApJ...660L...1D}. The GOYA Survey is a survey covering the Groth Strip compiling photometry in four optical bands ($U$, $B$, $F606W$, and $F814W$) and in two near infrared ones ($J$ and $K_{s}$) with visual classifications available, reaching similar depths to the Rainbow data in similar bands \citep[$U\sim 25$, $B\sim 25.5$, $K\sim 21$ mag, see][]{2003ApJ...595...71C,2006ApJ...639..644E,2007RMxAC..29..165A,2008A&A...488.1167D}.

We have performed the sample selection starting from a $K$-band selected catalog in the field, reaching a limiting magnitude for 50\% detection efficiency at $K\sim 21$\,mag. Several cuts have been performed to the original catalogue to obtain a mass-limited red galaxy sample. Firstly, red galaxies are selected as detailed in \S\ref{Sec:RGSelection}. This selection determines the redshift interval of the study, as it is restricted to the redshifts where the obtained number of red galaxies is statistically significant (i.e., to $0.3\lesssim z\lesssim 1.5$, see \S\ref{Sec:ClassificationSED}). 

According to the $M_{K}$-$z$ distribution of the red galaxies sample, the faintest absolute magnitude for which the catalogue is complete in luminosity up to $z\sim 1.5$ corresponds to $M_{K,\mathrm{lim}}\sim -24$\,mag. According to the redshift evolution of the mass-to-light relation (assuming a Salpeter IMF) derived by \citet{2007A&A...476..137A} for a sample of quiescent bright galaxies, a red passive galaxy with this $K$-band absolute magnitude at $z\sim 1.5$ has a stellar mass of $M_{*,\mathrm{lim}}\sim 5\times 10^{10}\Msun$. Therefore, we have selected red galaxies with masses higher than $M_{*}\sim 5\times 10^{10}\Msun$ at each $z$ just considering all galaxies with $M_{K,\mathrm{cut}} (z) =  -23.3 - 0.45 z$ to account for their luminosity evolution. This luminosity cut is very similar to the one obtained by \citet{2007MNRAS.380..585C}. Analogously, the equivalent luminosity cut to obtain a complete red galaxy sample for $M_{*,\mathrm{lim}}=10^{11}\Msun$ would be: $M_K = -24-0.45 z$ (we will use this selection for comparing our results with those reported in other studies, see \S\ref{Sec:RGSelection}). The detection efficiency in the $K$-band drops below 0.9 for $m_K >20.5$\,mag \citep{2003ApJ...595...71C}, so we have checked that all galaxies in our mass-limited red sample exhibit apparent magnitudes brighter than this limit. 

We have used the color-color diagram shown in Fig.\,\ref{Fig:star-galaxy} to remove stars from the sample. It represents the $I-K_{s}$ vs.~$V-I$ distribution for all the sources in the mass-limited red galaxy sample. Stars typically exhibit NIR colors bluer than galaxies, so they populate the lower region in the diagram. Attending to this bimodality of star-galaxy colors, we have defined a color-color cut to isolate galaxies from stars (see the solid line in the figure). The marked points include the stars and compact objects in the sample. We have checked that the stars identified according to it include all the objects at lower redshifts that have been classified as "stars" in the morphological classification performed in \S\ref{Sec:Morphology} (they are marked in Fig.\,\ref{Fig:star-galaxy}). 
As the number of compact objects found in the sample is not statistically significant (see \S\ref{Sec:Morphology}), they have been excluded from this study. From an initial sample of 1809 sources from the original $K$-band selected catalogue, we finally have a mass-limited red galaxy sample of 257 systems at $0.3<z<1.5$.

Figure\,\ref{Fig:zspec-zphot} compares the photometric and the spectroscopic redshifts for the red galaxies in our mass-limited sample with spectroscopic redshift determinations from the DEEP2 survey \citep{2007ApJ...660L...1D}.  The average redshift uncertainties are similar for both bright and faint sources at all redshifts, being below $\Delta(z)/(1+z) \sim 0.03$ in the whole redshift interval under study.

\begin{figure*}
\begin{center}
\includegraphics[width=\textwidth,angle=0]{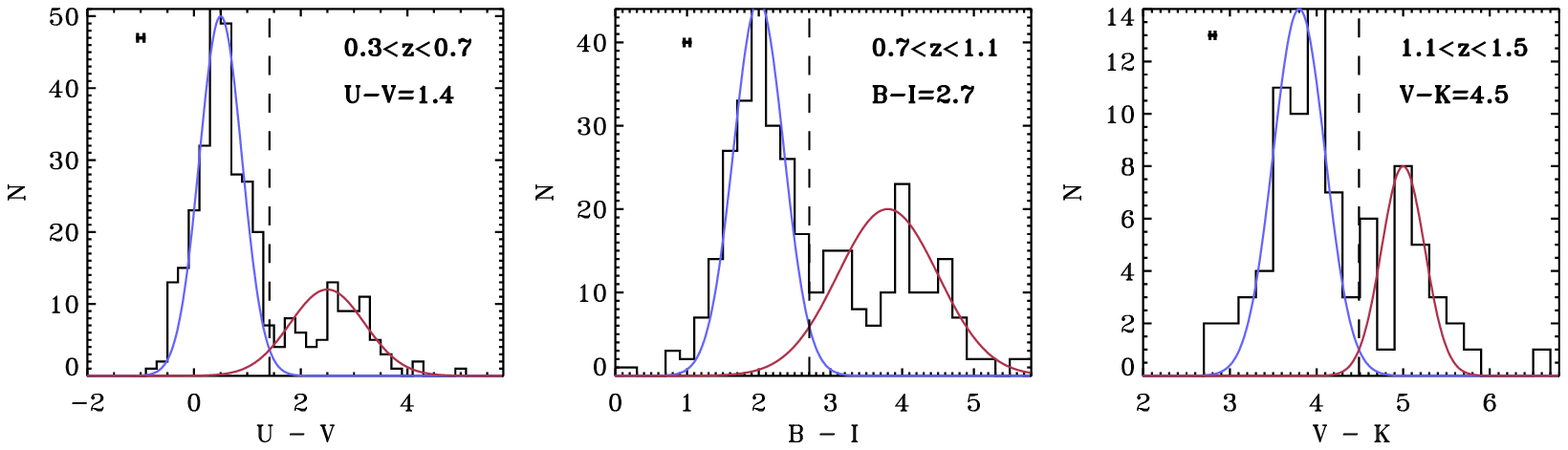}
\includegraphics[width=\textwidth,angle=0]{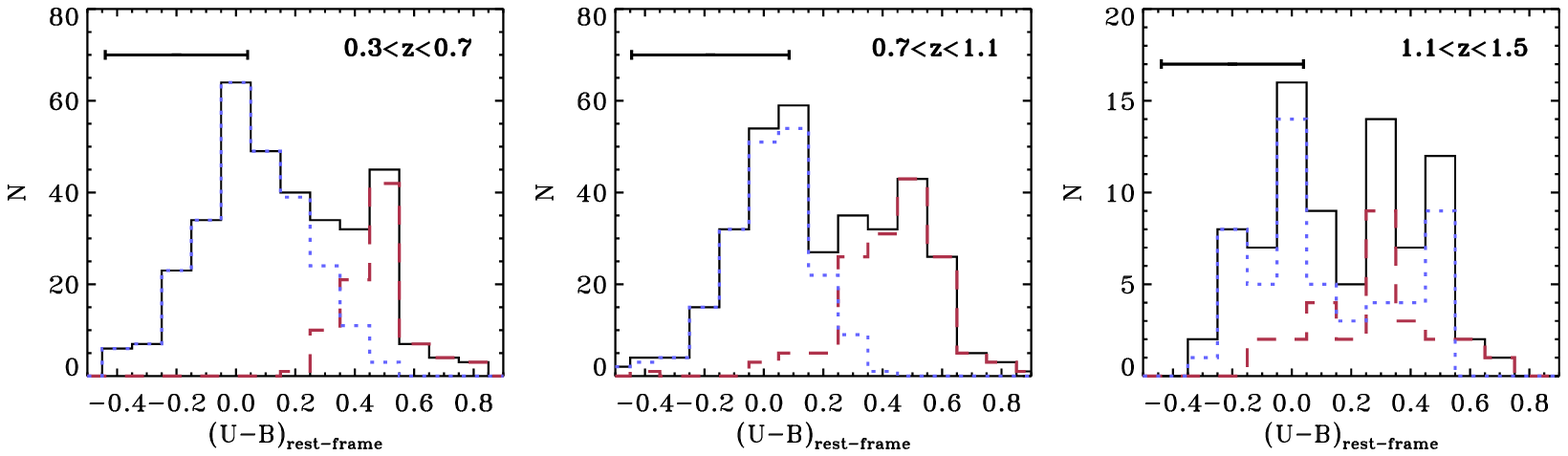}
\caption{Color histograms of all galaxies in the $K$-band selected catalogue in the three wide redshift bins under consideration ($0.3<z<0.7$,  $0.7<z<1.1$, and $1.1<z<1.5$). \emph{Top panels}: Histograms of the apparent color nearest to the rest-frame $U-B$ at each redshift interval. \emph{Red and blue solid lines}: Gaussian functions fitted to match the bimodal color distributions at each redshift. \emph{Vertical dashed lines in top panels}: Color cuts used to isolate red from blue galaxies in each redshift bin (see the text). \emph{Bottom panels}: Histograms of the rest-frame $U-B$ in the same redshift bins. The horizontal bar in the upper left corner of each frame represents the average color error of one galaxy in the sample in each case. Dashed red and dotted blue histograms in these panels show the distributions of red and blue galaxies, respectively, selected according to the criteria based on the apparent colors of the top panels. Red galaxies are selected on the basis of apparent colors just attending to the low uncertainties associated to them compared to the typical errors of the rest-frame $U-B$ colors at all redshifts.}
\label{Fig:simul}
\end{center}\end{figure*}

\section{Red galaxies selection}
\label{Sec:RGSelection}

Previous studies on red galaxies have been traditionally centered on the galaxies lying on top of the Red Sequence \citep[see, e.g.,][]{2000ApJ...541...95V,2006ApJ...647..853W,2007MNRAS.380..585C}. These studies derive a linear fit to the rest-frame $(U-B)_\mathrm{rest-frame}$-$M_B$ relation and consider all the objects within 1$\sigma$ scatter of this relation as red galaxies. Nevertheless, the majority of the galaxies that have migrated to the Red Sequence is expected to have spent part of their lives at nearby locations on the Green Valley, undergoing transitory stages of $\sim 1$\,Gyr in their evolution towards typical E-S0's (F07). Many observational and computational studies confirm that advanced stages of major mergers usually lay on the Red Sequence or at nearby locations on the Green Valley, because gas-rich encounters experience noticeable dust extinction due to the starbursts induced by the encounter \citep[see][]{1995AJ....110..129K,2001ApJ...550..212B,2003MNRAS.341...54K,2006MNRAS.370...74R,2008MNRAS.384..386C,2007MNRAS.380..585C,2009ApJ...693..112P,2010A&A...518A..61C,2008MNRAS.391.1137L,2010MNRAS.404..590L,2010MNRAS.404..575L}. Therefore, in this study we have included the systems both lying on the Red Sequence and on nearby positions on the Green Valley into our red galaxy sample, to also trace these transitory evolutionary stages towards the Red Sequence \citep[see more references in][]{2010A&A...519A..55E}.

\subsection{Color cuts for red galaxies selection}
\label{Sec:Colorcuts}

Following previous studies, we have selected red galaxies according to their rest-frame $U-B$ color \citep[see, e.g., F07;][]{2008ApJ...682..896K}. This color traces the 4000\,\AA\ break up, a spectral feature characteristic of evolved stellar populations. However, the rest-frame $U-B$ estimates derived from the present dataset exhibit so high uncertainties due to photometry and redshift-determination errors that we have used instead the observed color that samples the rest-frame $U-B$ more closely at the center of the three wide redshift bins covering the complete redshift range ($0.3<z<0.7$, $0.7<z<1.1$, and $1.1<z<1.5$).

The top panels of Fig.\,\ref{Fig:simul} show the color distribution of all galaxies in our $K$-band selected catalogue in the colors that sample more closely the rest-frame $U-B$ at these three redshift bins ($U-V$, $B-I$, and $V-K$ for the low, middle, and high  redshift intervals, respectively). Typical color errors of the galaxy sample are indicated with a horizontal bar at the left top corner of each frame. The figure shows the well-known bimodal distribution of galaxies into red and blue populations at all redshifts up to $z\sim 2$ (see references in \S\ref{Sec:introduction}). These bimodal distributions have been modelled as the addition of two Gaussian functions (also plotted in the figure). 

In order to include the galaxies located at nearby positions on the Red Sequence in the red sample (and not just the galaxies lying within 1$\sigma$ scatter of the red peak), we have adopted the colors at which both Gaussian distributions cross as the color cuts for isolating red from blue galaxies at each redshift bin. The resulting cuts are $(U-V)=1.4$\,mag for $0.3<z<0.7$, $(B-I)=2.7$\,mag for $0.7<z<1.1$, and $(V-K)=4.5$\,mag for $1.1<z<1.5$. 

In the bottom panels of Fig.\,\ref{Fig:simul}, we show the histograms of the rest-frame $U-B$ color in the same redshift bins. The dashed red and dotted blue histograms show the distribution of red and blue galaxies selected according to the criteria based on the apparent colors commented above. In $1.1<z<1.5$ redshift range, the red and blue galaxy distributions appear quite mixed, probably due to the high errors associated to rest-frame $U-B$ colors (see the horizontal bar in the upper left corner of each frame). These high errors in the rest-frame $U-B$ color inhibited us to perform the red galaxies selection on the basis of rest-frame $U-B$ colors. 

Previous studies report a negligible dependence of the rest-frame $U-B$ color cut isolating the Red Sequence from the Blue Cloud with the galaxy mass for both the mass and redshift range considered in this study \citep[the $U-B$ color cut varies $\sim 0.3$\,mag at most, see][]{2009ApJ...694.1171T,2011ApJ...727...51N}. Therefore,  any dependence of our color cuts with mass (or, equivalently, with luminosity) has been overridden.

\begin{table*}
 \begin{minipage}{140mm}
  \caption{Simulated star formation histories characteristic of different galaxy types.\label{Tab:SFH}}
\begin{center}
  \begin{tabular}{llrcl}
  \hline
\multicolumn{1}{c}{Galaxy Type} & \multicolumn{1}{c}{SFR law} & \multicolumn{1}{c}{Metallicity} & \multicolumn{1}{c}{$\tau_{*, V}$ (mag)\footnote{$\tau_{*, V}$: Optical depth of the stellar population in the $V$-band due to dust extinction.}} & \multicolumn{1}{c}{$z_\mathrm{f}$}\\
\multicolumn{1}{c}{(1)} & \multicolumn{1}{c}{(2)} & \multicolumn{1}{c}{(3)} & \multicolumn{1}{c}{(4)}& \multicolumn{1}{c}{(5)}\\  \hline
\multicolumn{1}{l}{Elliptical} & Finite burst ($\tau =0.7$\,Gyr) &  0.02& 0.6 & 0.8,1.0,1.3,1.5,2.0,2.5,3.0 \\[0.1cm]
 \multirow{2}{*}{S0 and spirals} & Exponentially declining with& \multirow{2}{*}{0.02} & \multirow{2}{*}{0.6}   & \multirow{2}{*}{1.2,1.3,1.5,2.0,3.0}\\
 & $e_f=1$, 2, 3, 5, and 7\,Gyr\footnote{$e_f$: Star formation e-folding timescales in exponentially declining SFRs.}& &   & \\[0.1cm]
\multicolumn{1}{l}{Irregular} & Constant&  0.008& 0.0& Present at all z\\[0.1cm]
\multicolumn{1}{l}{Dust-reddened}  & \multirow{2}{*}{Constant} & \multirow{2}{*}{0.008} & \multirow{2}{*}{3.0} & \multirow{2}{*}{Present at all z}  \\
\multicolumn{1}{l}{star-forming galaxy} &  &  &  &  \\
\hline
\end{tabular}
\end{center}
\vspace{-0.5cm}
\end{minipage}
\end{table*}

\begin{figure*}
\includegraphics*[width=0.33\textwidth]{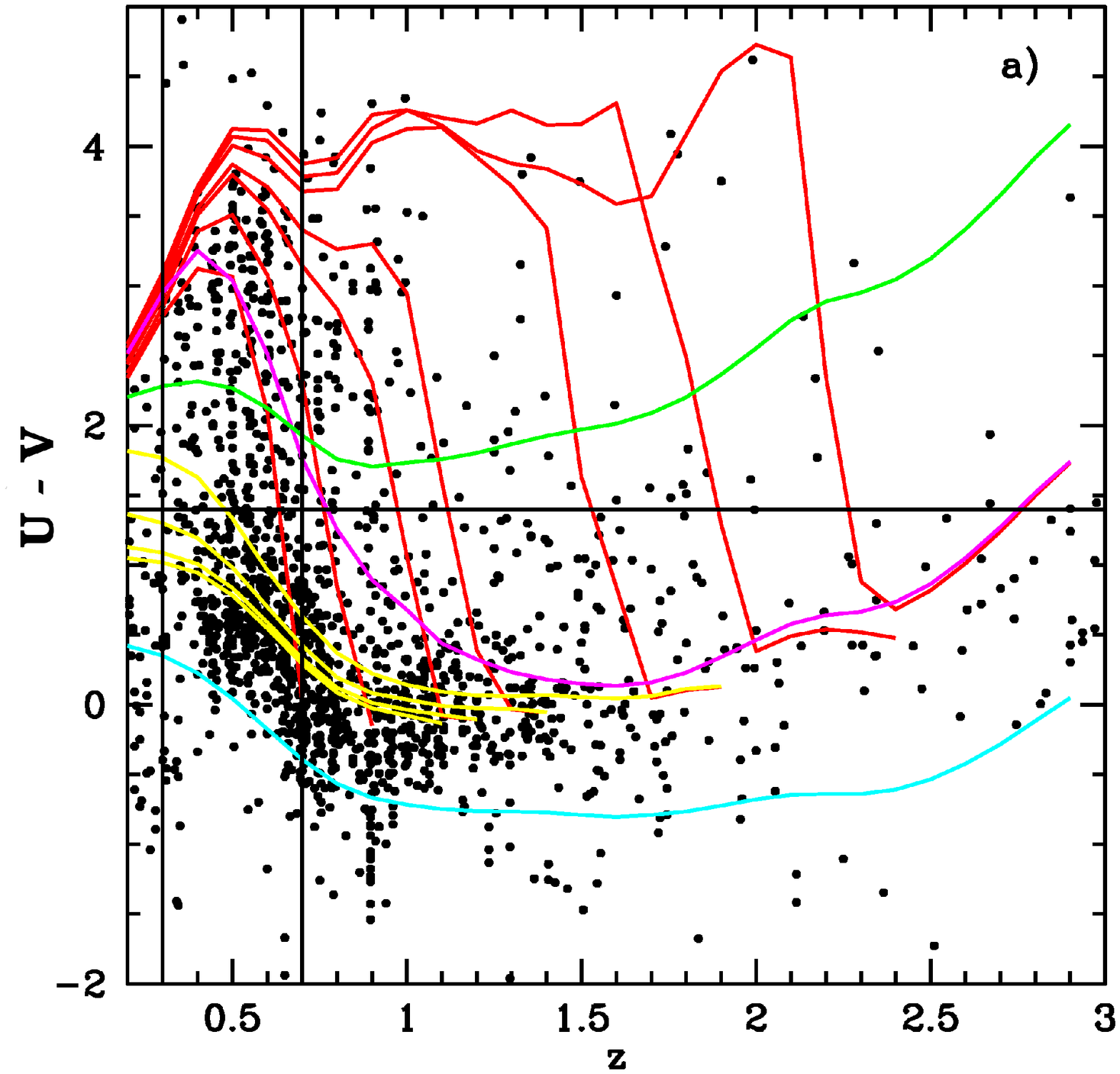}
\includegraphics*[width=0.33\textwidth]{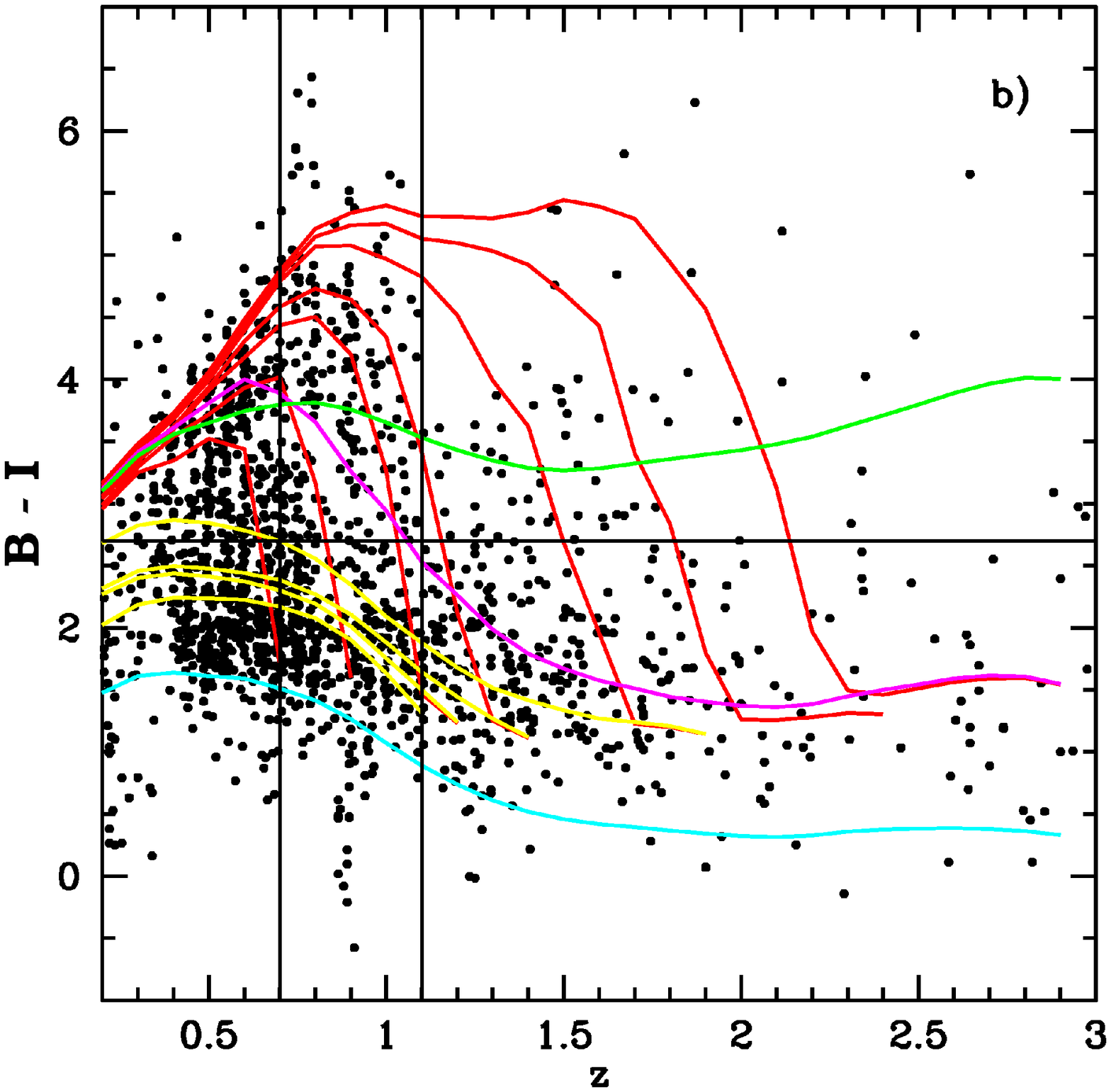}
\includegraphics*[width=0.33\textwidth]{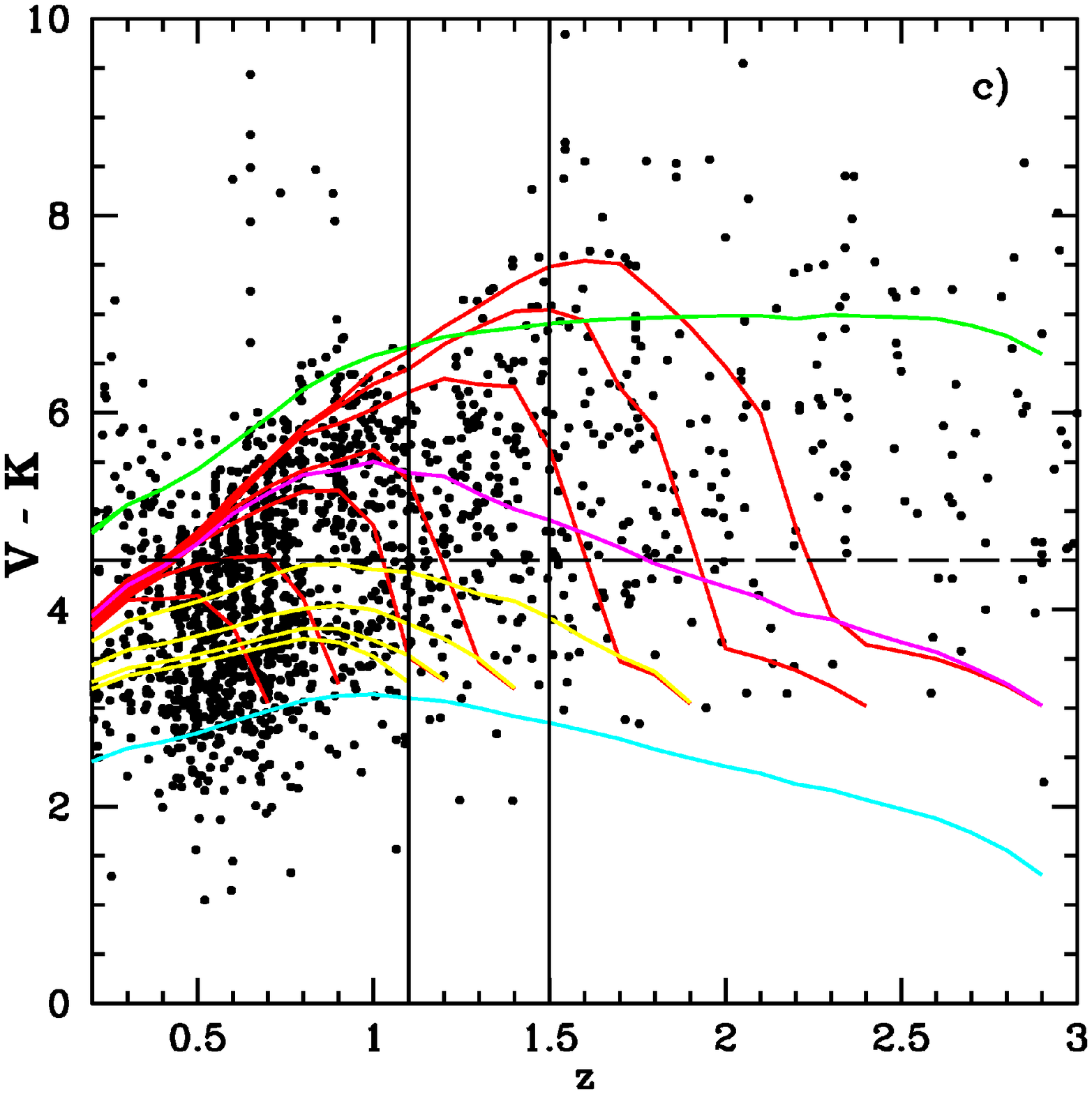}
\caption{Color-redshift distributions for all galaxies in the $K$-band selected catalogue, in the observed colors closest to the rest-frame $U-B$ color in the three wide redshift bins of the study. \emph{Panel a}: $U-V$ color (used for red galaxy selection at $0.3<z<0.7$). \emph{Panel b}: $B-I$ color (used for red galaxy selection at $0.7<z<1.1$). \emph{Panel c}: $V-K$ color (used for red galaxy selection at $1.1<z<1.5$). \emph{Dots}: Observational data. \emph{Lines}: Theoretical trends in the color-$z$ space for different galaxy types according to standard SFHs and physical properties of different galaxy types (more details in the text). \emph{Red}: Ellipticals with different $z_\mathrm{f}$. \emph{Magenta}: S0 galaxy with $z_\mathrm{f}=3$. \emph{Yellow}: Spirals with different $z_\mathrm{f}$. \emph{Blue}: Star-forming irregular. \emph{Green}: Dust-reddened star-forming galaxy with $\tau_{*, V}=3$. \emph{Horizontal black lines}: Color cuts defined to isolate red from blue galaxies, according to Fig.\,\ref{Fig:simul}. \emph{Vertical solid lines}: Limits of the redshift interval associated to the color plotted in each panel. Red galaxies in each redshift bin are those enclosed between the vertical lines and above the horizontal line marking the color cut.}
\label{Fig:colorvsz}
\end{figure*}

\subsection{Galaxy populations in the red galaxy sample}
\label{Sec:GalaxyPopulations}

We have analysed the kind of galaxy populations selected at each redshift interval with the color cuts defined in \S\ref{Sec:Colorcuts} to find out if the selection is homogeneous at each redshift bin. In Fig.\,\ref{Fig:colorvsz} we show the color-redshift distributions of all the galaxies in the $K$-band selected catalogue, for the color indices nearest to the rest-frame $U-B$ color in the three wide redshift bins considered in the study ($0.3<z<0.7$, $0.7<z<1.1$, and $1.1<z<1.5$). The color cuts to distinguish between red and blue galaxies at each redshift bin are marked in their corresponding panels with horizontal lines. The red galaxies selected at each redshift bin are those above the color cut and enclosed between the redshift limits associated to the color index in each panel. 

We have overplotted the theoretical evolution followed in each color-redshift plane by different galaxy types. These trends have been modelled using the IRAF package {\tt COSMOPACK\/} \citep{2003RMxAC..16..259B}, which uses the spectral energy distributions (SEDs) predicted by the stellar population synthesis models by \citet[]{2003MNRAS.344.1000B} to obtain the evolution of color indices with $z$. The galaxy types plotted in the figure include E's and spirals formed at different redshifts, a S0, a star-forming irregular, and a dust-reddened star-forming galaxy. Standard star-formation histories (SFHs), metallicities, and dust extinction values are assigned to each galaxy type, accordingly to observations  \citep{1989ApJ...344..685K,2005MNRAS.362...41G,2009ApJ...701.1965M}. The modelling parameters assumed for each simulated galaxy type are listed in Table\,\ref{Tab:SFH}. We must remark that we have adopted for the ellipticals a finite burst instead of the traditional single stellar population model. Although both SFRs provide similar colors and magnitudes evolution, the finite burst model can account for the short initial phases of star formation observed in these systems \citep{2009AJ....138..579K,2010A&A...515A...3H}.

Different formation redshifts have been considered for all the types. Nevertheless, as the redshift evolution of colors for S0's and their values were quite independent of $z_\mathrm{f}$, we have just plotted the S0 track for $z_\mathrm{f}=3$ in Fig.\,\ref{Fig:colorvsz} in benefit of clarity. As Irr's with different $z_\mathrm{f}$ are basically located below the spirals region in the plots, we only plot the case with $z_\mathrm{f}=3$ for the same reason. We have also simulated several cases of dust-reddened star-forming galaxies with different dust extinction levels. Their tracks basically overlap with those of the ellipticals in the plots, so we only show the model with $\tau_{*, V}=3$\,mag. 

 Although we have assumed different $z_\mathrm{f}$ for the formation of spirals, to assume the same epoch for their formation have a negligible effect on both their color evolution and location in the color-$z$ plane. They differ basically in their e-folding timescales, metallicities, and gas and dust contents. However, a similar conclusion cannot be derived for ellipticals. If we assume that all ellipticals have formed at an early epoch (e.g., $z_\mathrm{f}\geq 2$, as supported by some studies, see \S\ref{Sec:introduction}), the observed color-$z$ space of real red galaxies cannot be reproduced with the tracks modelled for $z_\mathrm{f}>2.0$, unless there is a large population of dust-reddened star-forming objects at all redshifts at $0.4<z<2$ \citep[][]{2006MNRAS.366..858K,2010A&A...519A..55E}. A wide range of values for the formation redshift of ellipticals reproduces more properly the observed color-$z$ distribution of red galaxies (see the mesh of red lines in each panel of the figure), a fact that does not exclude the existence of relevant populations of dust-reddened objects at different redshifts. 

The models for E's, S0's, and dust-reddened star-forming galaxies lie all above the color cuts in the three panels and that no spiral track enters the region of red galaxies in Fig.\,\ref{Fig:colorvsz}. This proves that the color cuts used in this study select quite homogeneous samples of galaxy populations in the whole redshift interval, basically galaxy types earlier than Sa and dust-reddened star-forming galaxies.

\begin{figure*}\begin{center}
\includegraphics[width=0.33\textwidth]{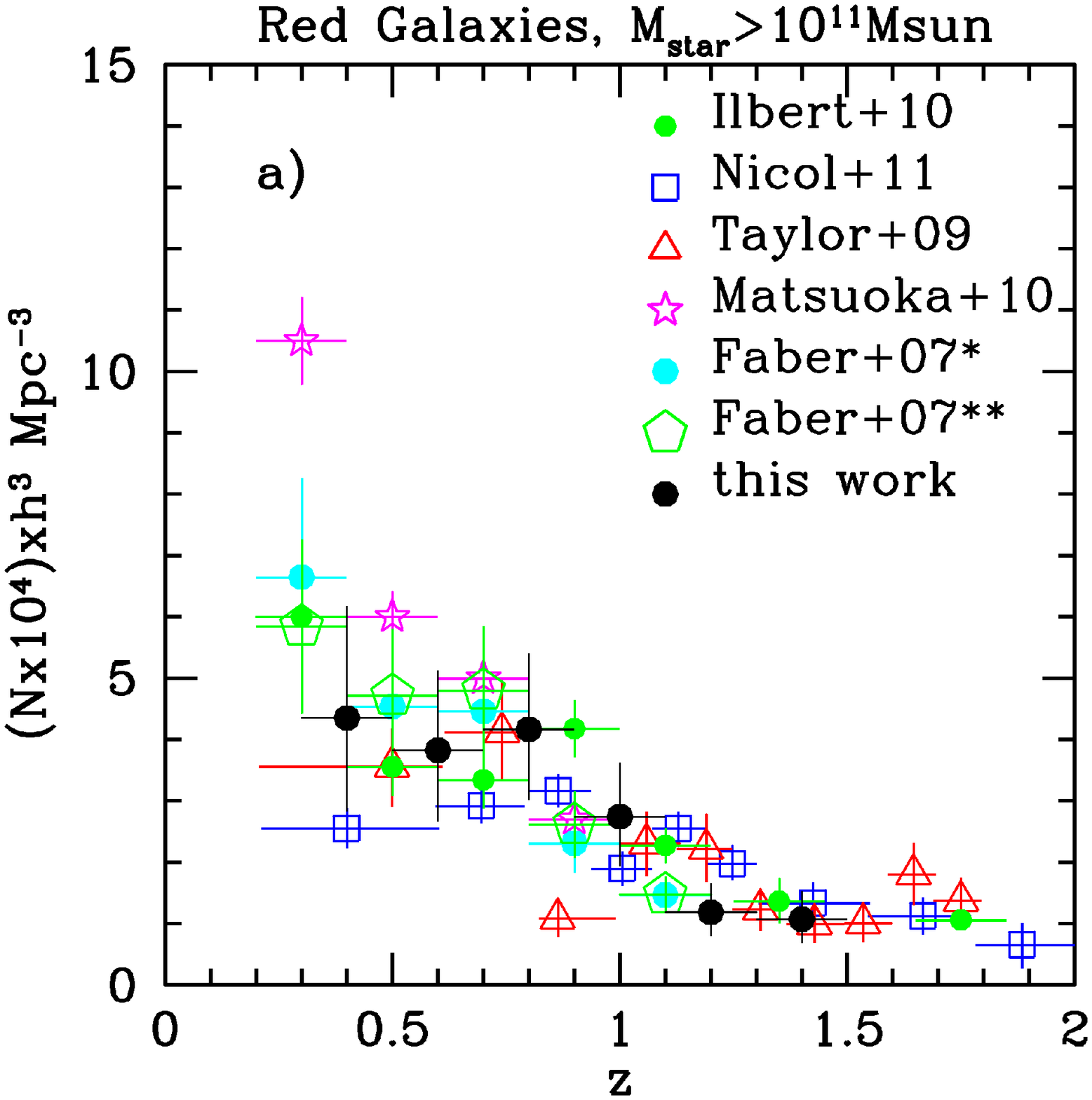}
\includegraphics[width=0.33\textwidth]{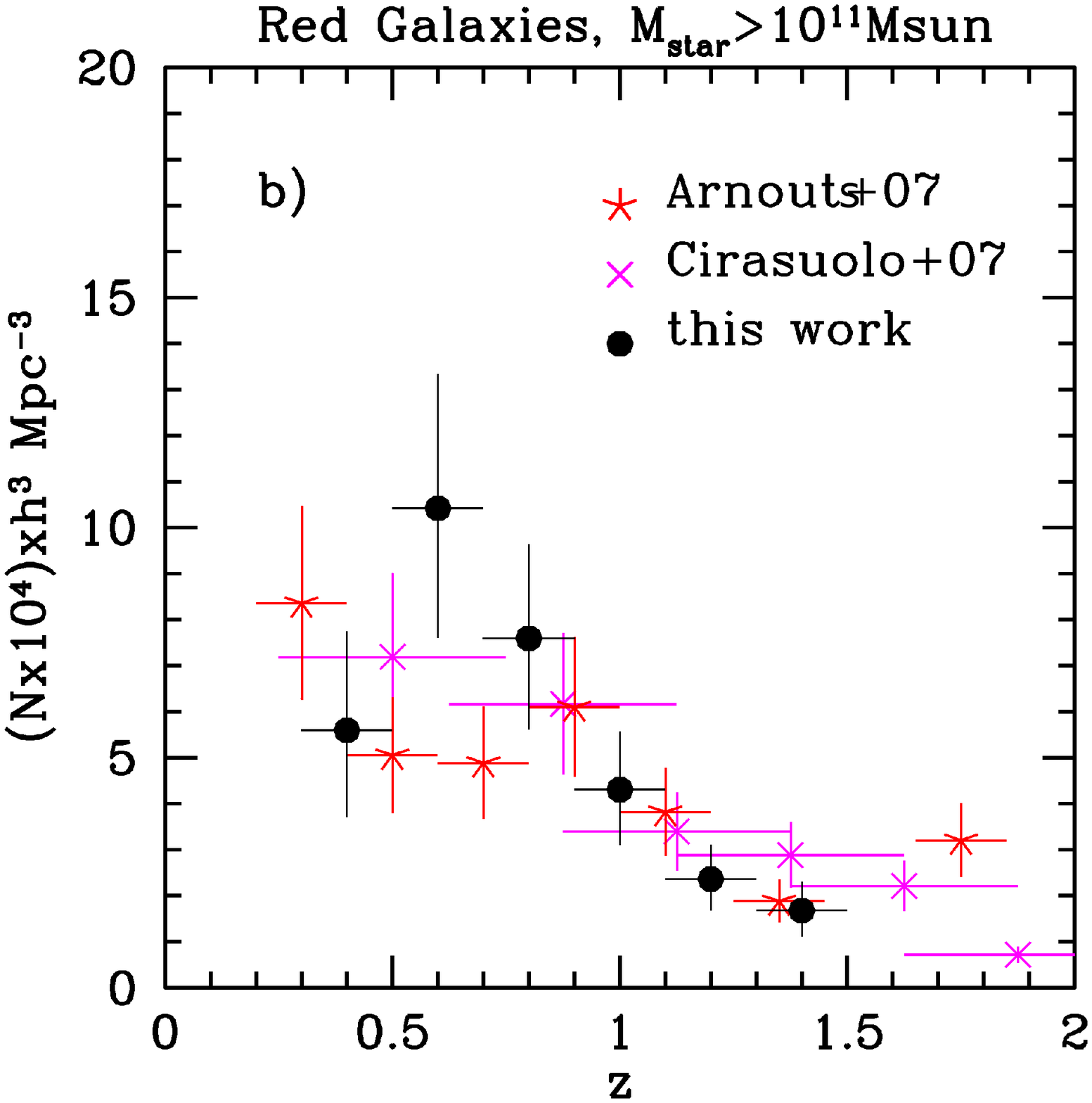}
\includegraphics[width=0.33\textwidth]{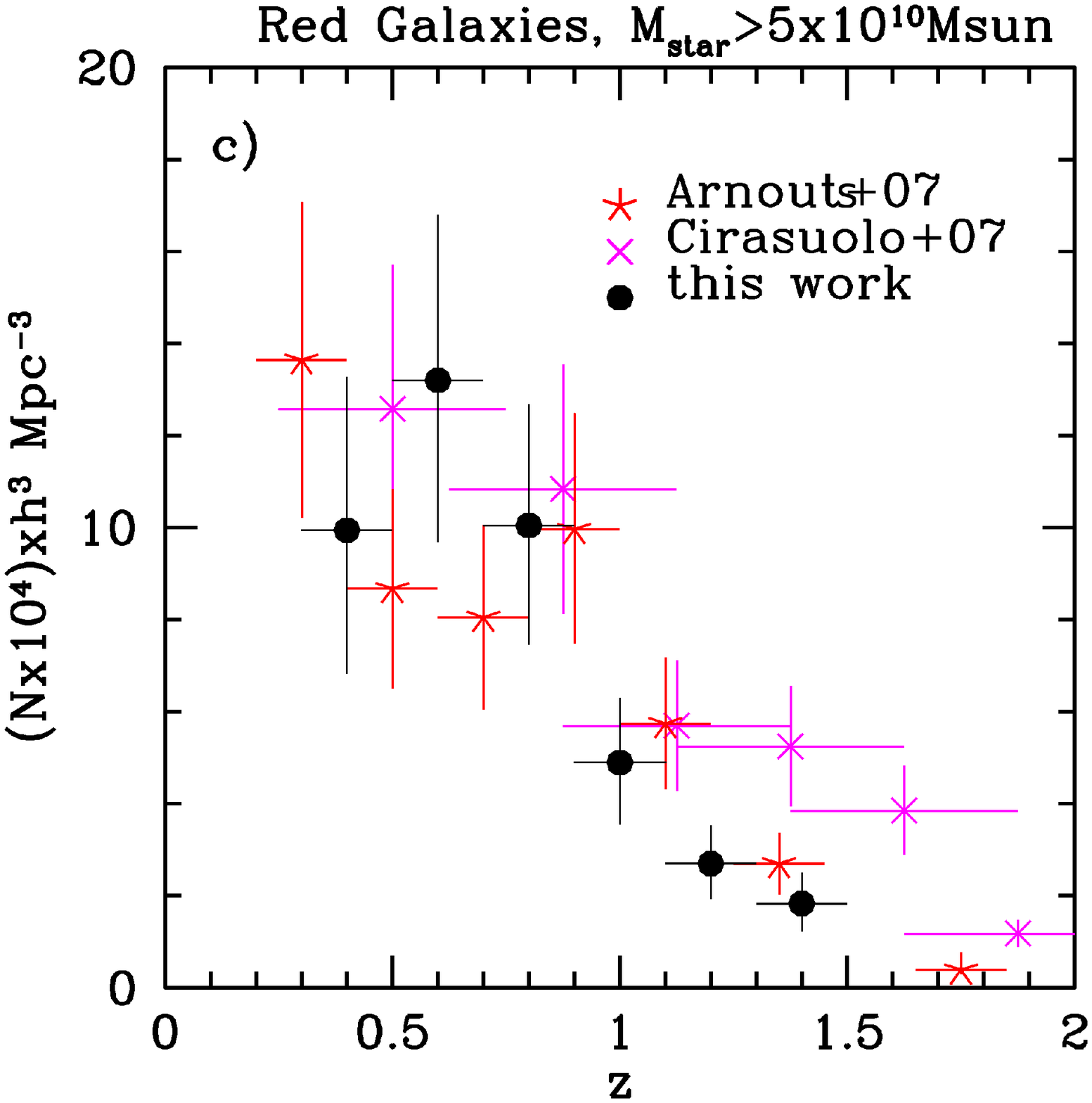}
\caption{Comparison of the redshift evolution of the number density of red galaxies obtained in this study with results from the literature for different mass and color selection criteria (see the text for details). Panels a and b plot the results for red galaxies with $\mathrm{M}_*>10^{11}\Msun$, but for different mass-to-light transformation relations. Panel c presents the results obtained with the same selection criteria as panel b, but for red galaxies with $\mathrm{M}_*>5\times 10^{10}\Msun$. All data plotted in the same panel use equivalent color and mass selection criteria. \emph{Filled black circles}: Results for our $K$-band selected data for each selection. \emph{Rest of symbols}: Results obtained by \citet{2007A&A...476..137A}, \citet{2007MNRAS.380..585C}, F07, \citet{2009ApJ...694.1171T}, \citet{2010ApJ...709..644I}, \citet{2010MNRAS.405..100M}, and \citet{2011ApJ...727...51N}. Plotted data assume $h\equiv H_0/100 = 0.7$ and a Salpeter initial mass function. Cosmic variance and sample incompleteness contribute to the large dispersion found among different studies at $z<0.4$.
\label{Fig:comparacion}}
\end{center}
\end{figure*}

\subsection{Comparison with other studies}
\label{Sec:Comparison}

The red galaxy selection made in this study cannot be directly compared to the red galaxy samples obtained by most studies in the literature because, first, we have included red galaxies adjacent to the Red Sequence to study objects at transitory stages of their evolution towards it (which is not usual, see \S\ref{Sec:introduction}), and secondly, we have estimated masses using the $\mathrm{M}_*/L_K$-z relation derived by \citet{2007A&A...476..137A} for different redshifts, whereas most authors use the $\mathrm{M}_*/L_V$-color relation derived by \citet{2001ApJ...550..212B} or an equivalent relation. Moreover, most studies report the number evolution of red galaxies for masses $\mathrm{M}_*>10^{11}\Msun$, instead of for masses $\mathrm{M}_*>5\times 10^{10}\Msun$ (as our case). In order to check out our results, we have made alternative red galaxy selections for $\mathrm{M}_*>10^{11}\Msun$, adopting the color cuts and/or the mass-to-light relation used by other authors.

The three panels of Fig.\,\ref{Fig:comparacion} compare the redshift evolution of the number density of red galaxies derived from our data with the results of different authors, for analogous mass and color selections in each case. In panel a, we have assumed the $U-B$ color evolution derived by \citet{2001ApJ...553...90V} to select red galaxies (following F07), and the masses are estimated using the $\mathrm{M}_*/L_V$-color relation by \citeauthor[][]{2001ApJ...550..212B}. Only red galaxies with $\mathrm{M}_*>10^{11}\Msun$ at each redshift are considered in this panel. Panel b of the figure also assumes the color cut by \citeauthor{2001ApJ...553...90V} for selecting red galaxies, but the mass estimates assume the $\mathrm{M}_*/L_K$ relation by \citeauthor{2007A&A...476..137A}, which includes evolutive corrections \citep[it is equivalent to the one derived by][]{2007MNRAS.380..585C}. The number densities of red galaxies shown in panel b are also for galaxies with  $\mathrm{M}_*>10^{11}\Msun$ at each redshift. Finally, panel c of the figure use the same selection criteria as panel b, but the number densities of red galaxies have been computed for $\mathrm{M}_*>5\times 10^{10}\Msun$. Note that the results of our color selection (including galaxies on the Red Sequence and at nearby locations) are not plotted in this figure (see \S\ref{Sec:Results} and Table\,\ref{Tab:densities}).

The data from \citet{2007A&A...476..137A}, \citet{2007MNRAS.380..585C}, and F07 have been obtained by integrating their red galaxy luminosity functions at each redshift for $\log(\mathcal{M}_*/\Msun)>11$. In this case, the absolute magnitudes have been transformed into stellar masses using the expression derived for local E-S0's by \citet{2006A&A...453L..29C}, considering the L-evolution of red galaxies derived by F07 and the redshift evolution of the $(B-K)$ color expected for E-S0's \citep[]{1998ApJ...501..578S,2003A&A...404..831D,2007A&A...476..137A}. AB magnitudes have been transformed to the Vega system in the $B$ and $K$ bands according to \citet{2007AJ....133..734B} transformations and considering galaxy colors derived for E-S0's by \citet{1995PASP..107..945F}.

The good agreement between our results and those from independent studies for similar selection criteria supports the reliability of our methodology and completeness of our nominal red sample (compare the black filled circles with the rest of studies in all panels of Fig.\,\ref{Fig:comparacion}). However, although our first data point in panels b and c in the figure is inside the cloud of points within errors, it does not follow the trend of the other authors. So, our data at this redshift are probably affected by volume and cosmic variance effects.

\section{Classification of red galaxy types}
\label{Sec:classification}

The hierarchical picture of galaxy formation predicts that massive E-S0's are the result of the most violent and massive merging histories in the Universe (see references in \S1 in EM10). To test this scenario, we need to distinguish between galaxies undergoing a major merger and normal E-S0's (see \S\ref{Sec:Tests}). Normal relaxed galaxies and major mergers differ basically in their structural distortion level. Major mergers also exhibit different global morphology and star formation enhancement depending on the gas content of the progenitors and the evolutionary stage of the encounter.

A gas-rich major merger is expected to turn into a dust-reddened star-forming disk with noticeable structural distortions at intermediate and advanced stages of the encounter, basically since the coalesce of the two galaxies into an unique galaxy body (this stage is known as the merging-nuclei phase). In earlier phases of the merger, the two galaxies can develop noticeable tidal tails and asymmetric structures, but the two bodies can still be distinguished and are not expected to suffer from enough dust reddening to lie nearby or on top of the Red Sequence. During the latest phase of the encounter (post-merger), the star formation is quenched and the remnant gets a more relaxed spheroidal structure until it transforms into a typical E-S0 \citep[see][]{2008MNRAS.384..386C,2008MNRAS.391.1137L,2010MNRAS.404..590L,2010MNRAS.404..575L}. Intermediate-to-late stages of gas-poor major mergers present a distorted spheroidal morphology and negligible levels of star formation, thus being quite red too \citep{2005AJ....130.2647V}. By the contrary, typical E-S0's present a spheroidal-dominated relaxed morphology, although they are also expected to be quite red due to their negligible star formation. Therefore, a gas-rich major merger is expected to be quite red from its merging-nuclei phase until its transformation into a typical E-S0, so we have traced major mergers once the two merging galaxies have merged into a unique remnant, because they are expected to be quite red in any case. Accounting for this, we have classified the galaxies in our red sample attending to their global morphology, structural distortion level, and star formation enhancement to distinguish among normal galaxies and intermediate-to-advanced phases of major mergers.

\begin{figure*}
\begin{center}
\includegraphics[width=\textwidth,angle=0]{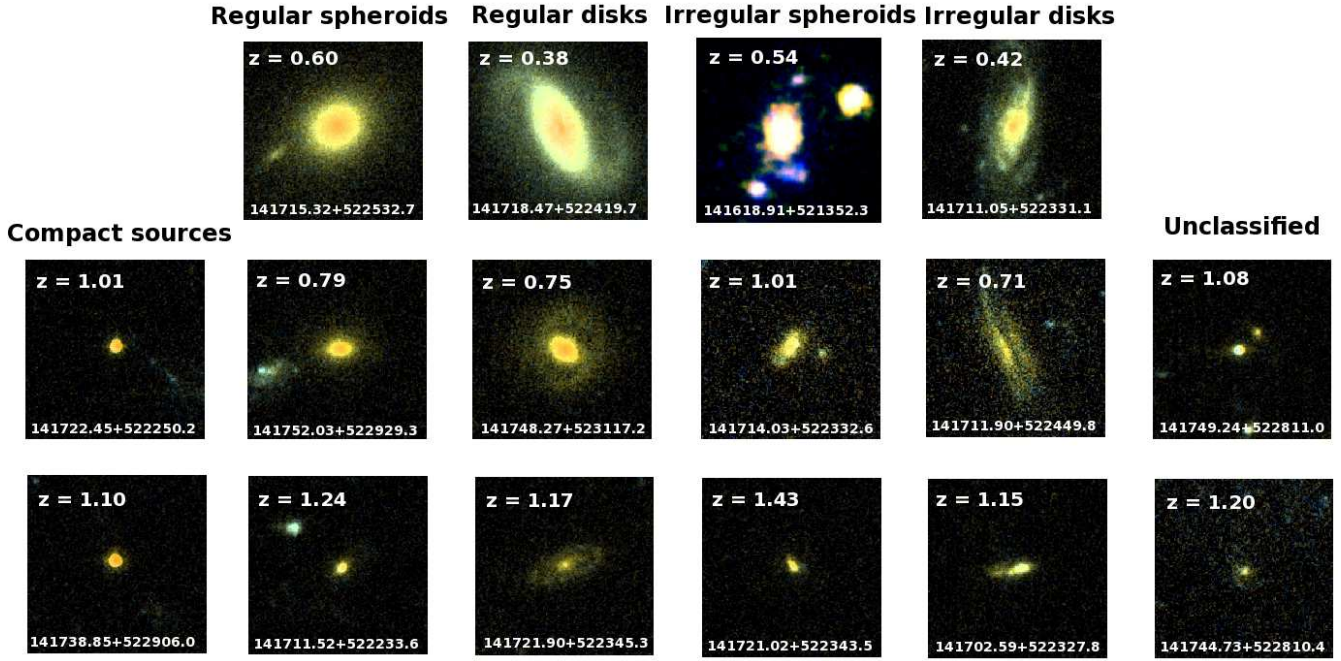}
\caption{False-color postage stamps of some red galaxies in our sample, obtained using the $V$ and $I$ bands. One example representative of each type is shown for each wide redshift bin used in the present study ($0.3<z<0.7$, $0.7<z<1.1$, and $1.1<z<1.5$). North is up, East is left. The frames correspond to a $5\arcsec\times 5\arcsec$ field-of-view, except for the irregular spheroid at $0.3<z<0.7$ (141618.91+521352.3), where a $10\arcsec\times 10\arcsec$ view is used to emphasize the interacting group to which this galaxy belongs (the galaxies at the North and at the Southern-East of the central object are located at the same redshift as the central object). No compact or unclassified objects are found at $0.3<z<0.7$. }
\label{Fig:stamps}
\end{center}\end{figure*}

\begin{figure*}
\begin{center}
\includegraphics[width=\textwidth,angle=0]{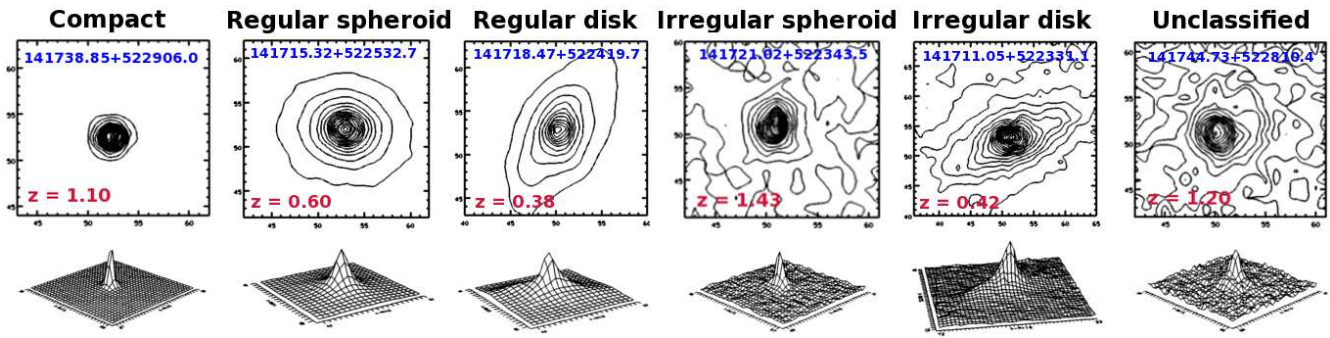}
\caption{$I$-band contour maps and surface plots of some red galaxies in Fig.\,\ref{Fig:stamps}. One galaxy representative of each type is shown ("C", "RS", "RD", "IS", "ID", "NC"). The axes in both the contour and surface maps are in CCD pixels.}
\label{Fig:stampscon}
\end{center}\end{figure*}

\subsection{Classification according to global morphology and distortion level}
\label{Sec:Morphology}

Despite the definition of very efficient quantitative morphological indices \citep{2008ApJ...672..177L}, visual inspection is still the most trustworthy method to classify galaxies morphologically \citep{2007MNRAS.382.1415S,2009ApJ...697.1971J,2010MNRAS.401.1043D}. The red galaxies in our sample have been thus classified visually by three co-authors. The comparison of the visual results with those obtained through quantitative classification methods has proven the reliability and robustness of the visual classification (see \S\ref{Sec:Robustness}).

No preconceived or classical morphological types have been used for the visual classification, as the emergence of the classical Hubble types seems to occur at $z\sim 1.0$-1.5 \citep{1996ApJ...472L..13O,2005ApJ...631..101P,2006ApJ...652..963R}. The visual types of red galaxies have been defined attending to the global characteristics exhibited by these galaxies in the sample. We find that red galaxies in our sample can be grouped morphologically attending to two aspects: 1) their structural distortion, and 2) its disk- or spheroid-dominated morphology. Six major exclusive classes have been identified according to the previous two criteria (see Fig.\,\ref{Fig:stamps}):

\begin{enumerate}
\item Compact galaxies (C).- Galaxies exhibiting compact morphologies, according to the seeing of the $I$-band images. No spatial information is available for them. The number of these objects in the sample is negligible, so we will not consider them henceforth.
\item Regular Spheroids (RS).- Galaxies with regular isophotes and dominated by a central spheroidal component. We remark that our RS's do not correspond to the spheroidal type defined by \citet{2012ApJS..198....2K}. This class basically groups E-S0 galaxies.
\item Regular Disks (RD).- Galaxies with regular isophotes and dominated by a disk component. Dust-reddened, typical spiral Hubble types correspond to this class.
\item Irregular Spheroids (IS).- Galaxies with irregular isophotes in the whole galaxy body but dominated by an spheroidal component.
\item Irregular Disks (ID).- Galaxies with irregular isophotes in the whole galaxy body and dominated by a disk.
\item Non-Classified or unclassified objects (NC).- Galaxies that cannot be classified into any one of the previous classes, because of their faintness or noise. The number of these objects is also negligible, so we will not consider them from now on.
\end{enumerate}

We have used surface brightness isophotes and surface maps to identify structural or morphological features that are more noticeable in these maps than in normal images (see Fig.\,\ref{Fig:stampscon}). We have also adopted the following additional classification rules. As commented above, we have traced exclusively advanced stages of major mergers, i.e., once both galaxies have merged into a unique body. Therefore, we have assigned a certain type to a red galaxy regardless of its environment, i.e., independently of whether it has close neighbours or not. We have considered as irregular galaxies only those systems that exhibit a noticeable distortion level in its \emph{whole} body. This implies that a galaxy in an interacting pair is identified as an independent regular galaxy, because its central body does not exhibit noticeable distortions yet, despite of the existence of significant tidal features in its outskirts. We remark that we are excluding from the irregular class the early phases of mergers in which the interacting galaxies have not merged into one body yet. The reason is because the tests for the hierarchical scenario of E-S0 formation described in \S\ref{Sec:Tests} are based in the tracing of these specific advanced stages of major mergers (and not of earlier phases). 

Minor mergers imprint less significant distortions than major mergers \citep{2006A&A...457...91E,2011A&A...533A.104E,2008MNRAS.391.1137L}, but they can also redden the galaxy enough to locate it at neighbouring regions of the Red Sequence temporarily \citep{2010MNRAS.404..590L,2010MNRAS.404..575L}. We have avoided the inclusion of minor mergers into the irregular class by considering exclusively morphological features typical of major mergers to classify a system as irregular, following \citet{2009ApJ...697.1971J}, such as the existence of multiple nuclei of similar luminosity in the body, the existence of equal-length tidal tails, or "train-wreck" morphologies. So, our red galaxy sample contain some minor mergers, but they are included into the regular class.

Our classification criteria try to minimize the effects of the spatial resolution loss and cosmological dimming inherent to the rise of redshift, which soften the luminosity distribution of galaxies, losing the physical regions of fainter surface brightness. Therefore, we have defined our galaxy types attending to features that affect to the whole galaxy body and that are poorly affected by these effects for the considered redshifts. Moreover, the classification has been done in the reddest visual band available with the best spatial resolution ($I$-band from HST/WFPC2). This band samples the rest-frame visual spectrum in the whole redshift range under study (from $I$ at low redshifts to $B$ at $z\sim 1.5$). Galaxies usually exhibit more distorted and clumpy morphologies in the rest-frame UV than in optical or NIR bands, but the morphology of a system is very similar in all rest-frame optical bands \citep{2000ApJS..131..441K,2009ApJ...697.1971J}, so we expect our visual classifications to not be biased towards irregular types at any redshift in our study. Simulations of the cosmological effects in the apparent $I$-band morphology of our systems have demonstrated the robustness of our visual classification against this effect (see \S\ref{Sec:TestVisual2}). In general, the three independent classifiers agreed in $\sim90$\% of the classifications at each redshift bin, a fact that strongly supports the robustness of the visual classification.

\begin{figure*}
\begin{center}
\includegraphics[width=0.32\textwidth,angle=0]{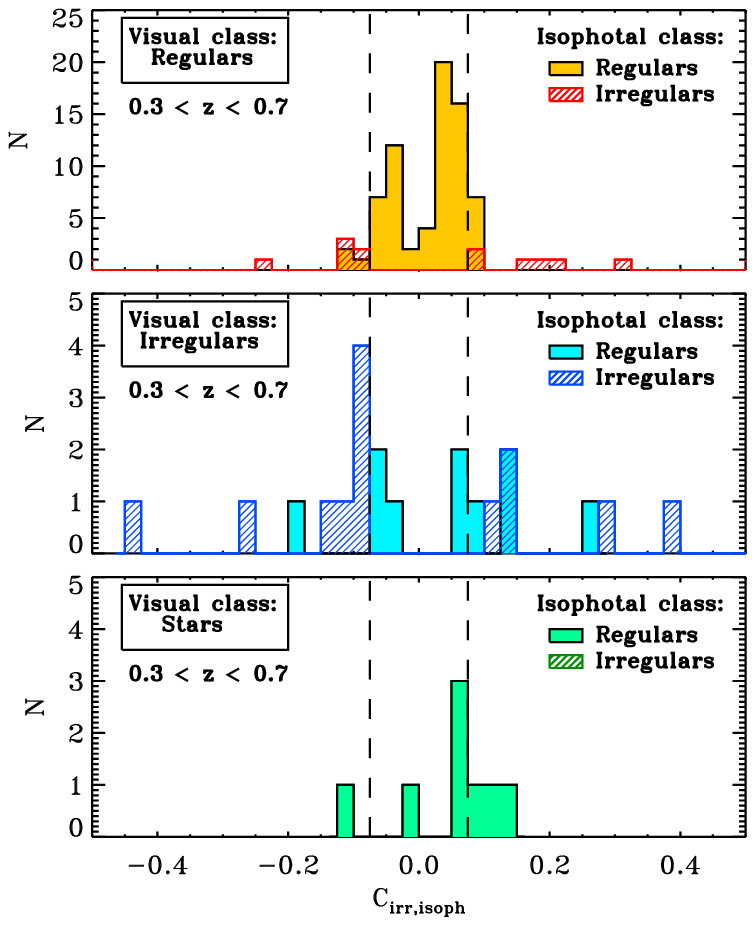}
\includegraphics[width=0.32\textwidth,angle=0]{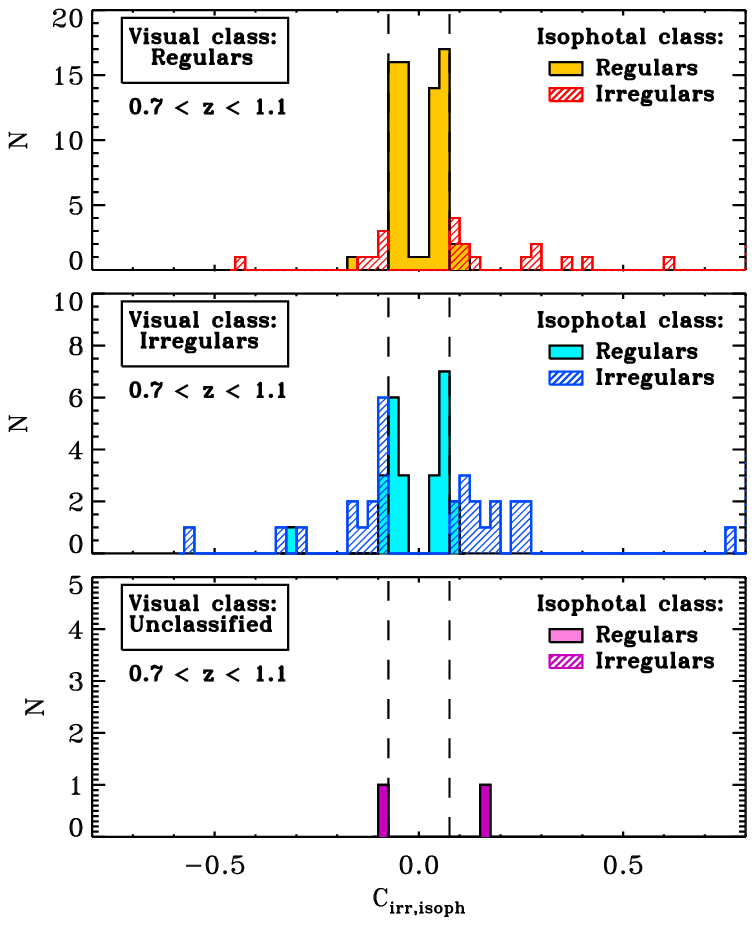}
\includegraphics[width=0.32\textwidth,angle=0]{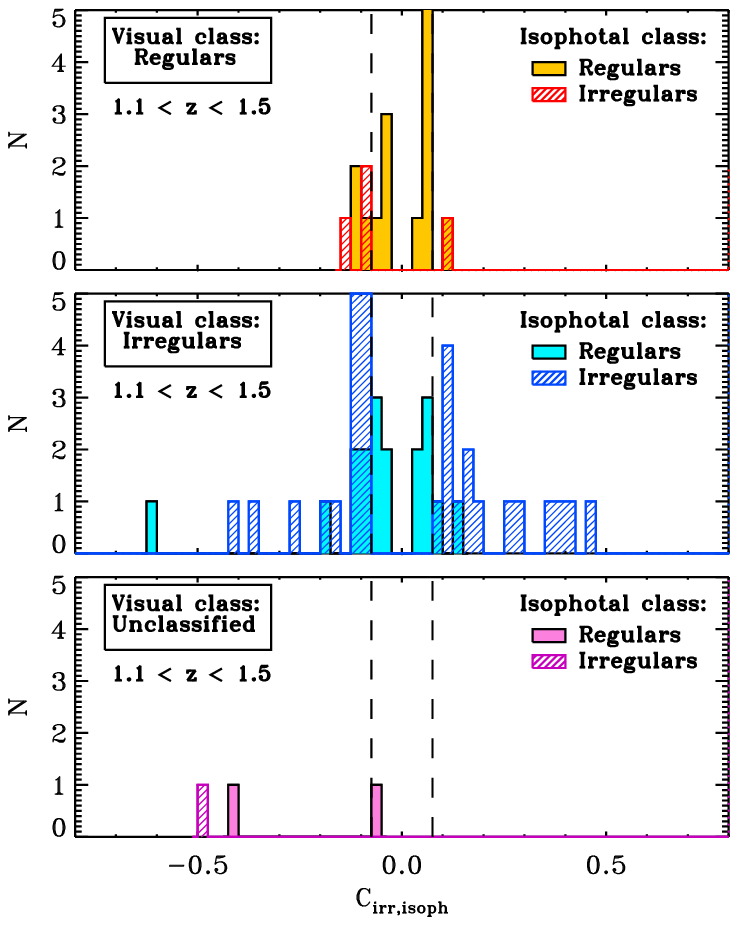}
\caption{Comparison of visual and quantitative morphological classifications into regular/irregular classes. The histograms of the galaxies corresponding to a given visual class are shown as a function of the quantitative irregularity parameter $C_\mathrm{irr,isoph}$ defined in \S\ref{Sec:TestVisual1}, for the three wide redshift bins of the study. \emph{Left panels}: For $0.3<z<0.7$. \emph{Middle panels}: For $0.7<z<1.1$. \emph{Right panels}: For $1.1<z<1.5$. Visual regulars, irregulars, and stars or unclassified objects are plotted from the top to the bottom panels of each column, respectively. The galaxies classified as regular/irregular according to the quantitative method are indicated in each panel. \emph{Vertical dashed lines}: Limiting values of $C_\mathrm{irr,isoph}$ to classify a galaxy as regular ($|C_\mathrm{irr,isoph}|=0.05$). Galaxies with $|C_\mathrm{irr,isoph}|<0.05$ are regular. However, $|C_\mathrm{irr,isoph}|$ values above this limit do not necessarily imply irregular morphology (see the cases in the text). Visual and quantitative classifications of irregularity agree in $\sim 75$\% of the cases at all redshifts on average, supporting the robustness of the visual method.}
\label{Fig:IrregularClassification}
\end{center}\end{figure*}

\subsection{Tests to the robustness of the visual morphological classification}
\label{Sec:Robustness}

We have carried out two kind of quantitative tests to check the robustness of the visual classifications:

\begin{enumerate}
 \item Tests on the reliability of the visual distinction of spheroid/disk-dominated and irregular/regular classes, based on the quantification of the light concentrations and the deviations of the isophotes from perfect ellipses in the whole galaxy body. 

 \item Tests on the cosmological effects on the visual morphology in the observed $I$-band (resolution loss, 
cosmological dimming, and change of the observed rest-frame band with redshift). 
\end{enumerate}

\subsubsection{Reliability of the visual classifications}
\label{Sec:TestVisual1}

In order to check our visual classification, we have computed concentration and asymmetry indices as defined by \citet{2003ApJS..147....1C}, which are quantitative indices closely related to our visual classes by definition (irregular galaxies should exhibit high asymmetries, while spheroid-dominated galaxies must have high concentrations). We find that galaxies visually classified as RS's show high concentration indices in general ($\sim 90$\%). The irregulars (IS's and ID's) present a wide spread of concentration values, consistently with their merger-related nature. 

Nevertheless, we also find that the galaxies classified visually as irregulars do not exhibit any correlations with the asymmetry index, contrary to expectations. This is probably because the asymmetry index estimates are extremely affected by background substructures, thus requiring high signal-to-noise data to be reliable \citep[$S/N>100$, see][]{2000ApJ...529..886C,2003ApJS..147....1C,2009MNRAS.394.1956C}. But the galaxies in our red sample have $S/N \sim 40-50$ at most (as $I$-band magnitude errors are $\sim 0.02-0.03$\,mag typically), making our asymmetry estimates quite uncertain. Moreover, asymmetry indices are sensitive to environmental influences in the galaxy outskirts. This means that some galaxies identified as regulars according to our criteria (because they do not exhibit noticeable distortions in its whole body) may have a high asymmetry index because of tidal features in the outer parts. This obviously smudges the correlation between visual irregularity and computed asymmetries. 

We have adapted the method by \citet{1993ApJ...418...72Z} to quantify the irregularity level of galaxy morphology. These authors developed a procedure to classify an elliptical galaxy as regular or irregular, attending to the distortion level of their isophotes with respect to perfect ellipses. They fitted ellipses to the isophotes of each elliptical to obtain the radial profiles of the coefficients $a_3$, $b_3$, $a_4$, and $b_4$ of their Fourier expansion series. The peak value of each Fourier coefficient was identified along the radial profile. These authors considered the following criteria to distinguish among regular and irregular galaxies:

\begin{enumerate}
 \item If all the peaks of the coefficients were small, this meant that the isophotes exhibited small deviations 
   from perfect ellipses. Therefore, these galaxies could be considered as regulars.
 \item If one of them was not small, then they differentiated between two possible cases:
    \begin{itemize}
     \item If the maximum value of the peaks of all coefficients did not correspond to $b_4$, then the isophotes deviated noticeably from ellipses, meaning that the galaxy was irregular.
     \item If the peak of $b_4$ was the maximum among all the peaks, its classification depended on its trend with the radial position in the galaxy. If the profile of this coefficient changed from one type to another within 1.5 effective radius of the galaxy, then the galaxy was irregular. If not, it just implied that the galaxy was boxy or disky (depending on the sign of $b_4$), but the galaxy morphology could be considered as regular.
    \end{itemize}
\end{enumerate}

We can adopt this method for our galaxies, as we are considering irregularities that must affect to the galaxy as a whole. We have limited the analysis in each red galaxy to the isophotes with a mean signal higher than 1.5 times the standard deviation of the sky signal per pixel. We have used the IRAF task \texttt{ellipse} for fitting ellipses to the isophotes and for getting the third and fourth order coefficients of their Fourier expansion series. We have identified the peaks of each coefficient in the galaxy radial profile. 

The task \texttt{ellipse} uses a normalization to the surface brightness of the isophote that directly measure the deviations of the isophote from perfect ellipticity. According to de \citet{1990AJ....100.1091P}, these deviations can be considered negligible if they are $< 5$\%, a value that can be translated directly to a value of 0.05 in these coefficients. Therefore, we have adopted this limit as the critical value to distinguish between small and high values. We have defined the irregularity index $C_\mathrm{irr,isoph}$ as the peak value of maximum absolute value among the peaks of the four Fourier coefficients. Therefore, according to \citet{1993ApJ...418...72Z}, any galaxy that has $|C_\mathrm{irr,isoph}|<0.05$ is regular. If not, it is irregular, except if $C_\mathrm{irr,isoph}$ corresponds to the $b_4$ coefficient and this coefficient does not change between $|b_4|>0.5$ and $|b_4|<0.5$ values or viceversa along its radial profile. 

We compare the results of the visual and quantitative classifications concerning the irregularity level of our red galaxies in Fig.\,\ref{Fig:IrregularClassification} for each wide redshift bin. The percentage of agreement between the visual and quantitative classifications into the regular type is $\sim 77$\% (decreasing from 83\% to 70\% from low to high redshifts). This percentage is slightly lower in the irregular type: $\sim $66\% (rising from 58\% to 74\% from low to high redshifts). The miss-classifications between both methods are $\sim $23\% for visually-identified regular galaxies (rising the confussion percentage from 17\% to 30\% from low to high z), and $\sim $34\% for visual irregulars (dropping from 42\% at low z to 26\% at high z). In general, both procedures coincide in $\sim 78$\% of the galaxy classifications at $0.3<z<0.7$, in $\sim 68$\% of the classifications at $0.7<z<1.$1, and in $\sim 73$\% at $1.1<z<1.5$. 

It is important to remark that only isophotal data above 1.5 times the sky standard deviation have been considered to quantify the irregularity of the galaxy in the quantitative method, i.e., it analyses isophotes of $S/N> 1.5$ at all redshifts. This implies that the quantitative method is not biased towards more irregular types at high redshifts, as it is limited to the isophotes with enough $S/N$ at all redshifts. Obviously, $C_\mathrm{irr,isoph}$ is derived from an intrinsic physical region in the galaxies smaller at high redshift than at low redshift, just because of cosmological effects. But, as commented above, we consider as irregular galaxies only the stages of advanced major mergers, which imply a noticeable distortion level in its whole body. The effects of cosmological dimming and resolution loss on the classification are analysed in \S\ref{Sec:TestVisual2}.

In conclusion, this test proves the robustness of the visual classifications into regular and irregular types at all redshifts.

\subsubsection{Robustness of the observed morphology against cosmological effects}
\label{Sec:TestVisual2}

In order to find out how the loss of spatial resolution, the cosmological dimming, and the change of rest-frame band with redshift are affecting to our classification, we have simulated images of galaxies at different redshifts in the observed $I$-band. We have used \texttt{COSMOPACK} \citep{2003RMxAC..16..259B}, an IRAF package that transforms images of real galaxies to depict their appearance at a given redshift as observed with a given telescope, camera, and filter. The transformation includes K-corrections, change of observing band, repixelation to the scale of the observing system, convolution by the seeing, and noise from sky, detector, and dark current. 

Starting from the $I$-band image of a galaxy representative of one type at the lowest wide redshift bin ($0.3<z<0.7$), we have simulated the observed $I$-band image of the same object at the middle of the other two redshift bins ($z= 0.9$ and $z=1.3$), in order to check if its morphological classification changes as the object is placed at higher redshifts. We have assumed a typical SFH, metallicity, and formation redshift for the galaxy to assign a characteristic SED  at its original redshift. Once we obtain the simulated image at $z=0.9$ and $z=1.3$, we have classified them visually following the same procedure as with the original image. 

The surface maps at different redshifts show that the distinction between the spheroid/disk-dominated morphology is quite robust in the whole redshift interval under consideration. In general, we can distinguish the winged surfaces of disks in the surface maps at all redshifts (see Figs.\,\ref{Fig:stamps}-\ref{Fig:stampscon}). This means that the visual classification into spheroid/disk-dominated systems is quite robust against cosmological effects for the redshifts and magnitudes studied here.

Concerning the irregularity, the contour plots support the expected trend of external isophotes to appear more irregular as we move towards higher redshifts. But this effect should not affect strongly our results, as we derive our visual classification into regular/irregular classes having into account the morphology of the isophotes over the whole galaxy body, discarding the most external ones. This qualitative test indicates that cosmological effects are not expected to affect noticeably the classification into regular/irregular types either.

\begin{figure}
\begin{center}
\includegraphics[width=0.5\textwidth,angle=0]{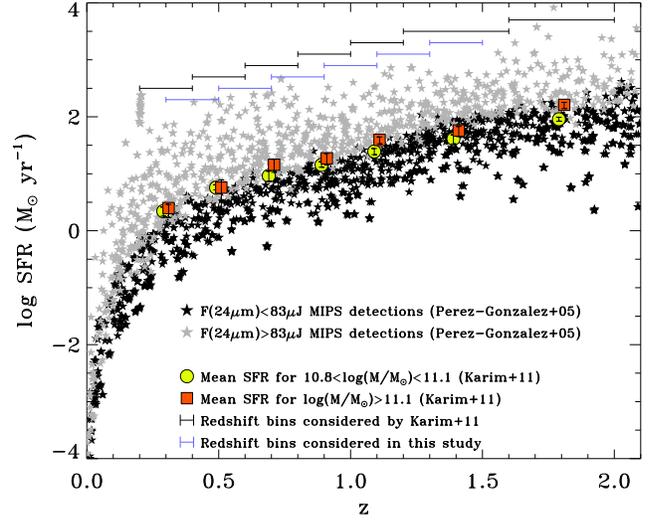}
\caption{SFR cuts selected to define the red galaxies with enhanced SFRs compared to the average SFR of the galaxy population, as a function of redshift. \emph{Red squares}: Redshift evolution of the SFR of galaxies with $\log(\mathcal{M}_*/\Msun) >11.1$ derived by \citet{2011ApJ...730...61K}. \emph{Yellow circles}: Redshift evolution of the SFR of galaxies with $10.8 < \log (\mathcal{M}_*/\Msun) < 11.1$ derived by the same authors. \emph{Grey stars}: Galaxy data from \citet{2005ApJ...630...82P} with $F(24\mu m)> 85$\,$\mu$J. \emph{Black stars}: Galaxy data from the same authors with $F(24\mu m)< 85$\,$\mu$J. This limiting flux in 24\,$\mu$m naturally isolates galaxies with high (enhanced) SFR compared to the average value of the whole galaxy population at each redshift from those with low SFR, for the mass range considered here.} 
\end{center}
\label{Fig:ksfr}
\end{figure}

\subsection{Classification according to star formation activity}
\label{Sec:ClassificationSED}

A typical problem of studies based on red galaxy samples is to disentangle dust-reddened star-forming galaxies from quiescent ones \citep[][]{2000MNRAS.317L..17P,2004ApJ...617..746D,2007ApJ...665..265F,2011MNRAS.412..591P}, because both galaxy populations are indistinguishable using only broad-band photometry at wavelengths $\lesssim 10$\,$\mu$m \citep{2006AJ....132.1405S}. In particular, the rest-frame $U-B$ color cannot differentiate between both red galaxy types adequately at $z>0.8$ (F07). Therefore, different selection techniques based on color indices, mid-IR data, or SED fitting have been developed to isolate both populations in red galaxy samples \citep{1999ApJ...518..533L,2003A&A...401...73W,2006A&A...455..879Z,2009ApJ...691.1879W}.

The mid-IR emission and the SFR of a galaxy are known to be tightly correlated \citep[][]{1998ARA&A..36..189K,2005ApJ...630...82P}. The higher sensitivities and spatial resolutions achieved by IR instruments in the last years have allowed the development of SFR indicators based on the emission of a galaxy in a single mid-IR band  \citep[]{2005ApJ...633..871C,2007ApJ...666..870C,2006ApJ...648..987P}. In particular, the 24\,$\mu$m band of the Multiband Imaging Photometer in the \emph{Spitzer} Space Telescope (MIPS) is found to be a good tracer of the infrared emission coming from the dust heated by star-forming stellar populations \citep{2006ApJ...650..835A}. 

In this study, we have identified galaxies with noticeable star formation compared to the average SFR exhibited by the galaxies with similar masses at the same redshift, because this is an evidence of that mechanisms different to passive evolution are triggering it (such as tidal interactions, mergers, gas infall, or stripping). The SFR of a galaxy changes noticeably with redshift due to the natural evolution of their stellar populations. The specific SFR has decayed with cosmic time as $\sim (1+z)^{n}$ since $z=3$, being $n=4.3$ for all galaxies and $n=3.5$ for star-forming sources \citep{2011ApJ...730...61K}. This means that we must have into account the intrinsic rise of SFR with redshift to define when a galaxy is forming stars more efficiently than the average of the whole galaxy population at each redshift. Therefore, we have used {\it Spitzer}/MIPS 24~$\mu$m data to discriminate between red galaxies with enhanced SFRs from those that have lower SFRs compared to the average SFR of the whole galaxy population at each redshift \citep[see also][]{2009ApJ...691.1879W}. A limiting flux in 24\,$\mu$m of $\sim 60$\,$\mu$Jy corresponds to the 5-$\sigma$ detection level on our MIPS 24~$\mu$m data \citep[][]{2011ApJS..193...13B}.

In Fig.\,\ref{Fig:ksfr}, we plot the $z$-evolution of the average SFR of galaxy populations with different mass ranges overlapping with ours (which is $\log (\mathcal{M}_*/\Msun)>10.7$), as derived by \citet{2011ApJ...730...61K} from a deep 3.6\,$\mu$m-selected sample in the COSMOS field. We have overplotted the location of the galaxies from \citet{2005ApJ...630...82P} in the diagram, differentiating those with emission fluxes in MIPS 24~$\mu$m above 85\,$\mu$J from those below it. Note that this limiting flux follows pretty well the redshift evolution of the average SFR of the whole galaxy population for our mass range, being a straightforward way for distinguishing galaxies with enhanced SFRs from those with low SFRs, compared to the average value of the galaxy population at each redshift. Therefore, we have considered red galaxies with a 24~$\mu$m MIPS flux above $85$\,$\mu$Jy as systems with enhanced star formation compared to the global galaxy population at the same redshift (High Star-Forming galaxies or HSF, hereafter), whereas the galaxies with a 24~$\mu$m flux below this limit are not substantially star-forming compared to it (so, they are Low Star-Forming ones or LSF, henceforth). 

\begin{table}
\begin{minipage}{0.5\textwidth}
\caption{Percentages of AGN contamination in the subsample of HSF red galaxies by morphological types.\label{Tab:agns}}
\begin{center}
\begin{tabular}{lcccccccc}
\hline
 \multirow{2}{*}{HSF}   & \multicolumn{2}{c}{$0.3<z<0.7$}   & & \multicolumn{2}{c}{$0.7<z<1.1$} & &  \multicolumn{2}{c}{$1.1<z<1.5$}  
\vspace{0.05cm}\\\cline{2-3}\cline{5-6}\cline{8-9}\vspace{-0.2cm}\\
type &\multicolumn{1}{c}{Pure$^\mathrm{a}$} & \multicolumn{1}{c}{Total$^\mathrm{b}$} 
& &\multicolumn{1}{c}{Pure} & \multicolumn{1}{c}{Total} & &\multicolumn{1}{c}{Pure} & \multicolumn{1}{c}{Total}  \\
 & (\%) & (\%) & & (\%) & (\%) & & (\%) & (\%) \\
\multicolumn{1}{c}{(1)} & \multicolumn{1}{c}{(2)} & \multicolumn{1}{c}{(3)} & & \multicolumn{1}{c}{(4)} & \multicolumn{1}{c}{(5)}& & \multicolumn{1}{c}{(6)} & \multicolumn{1}{c}{(7)}\\\hline
RS  &  0 & 0 & & 22 & 22 & & 0 & 0\\
RD  &  0 & 0 & & 0 & 0 & & 0 & 0\\
IS   &  0 & 0 & & 13 & 38 & & 0 & 13\\
ID   &  0 & 0 & & 0 & 0 & & 0 & 0\vspace{0.12cm}\\
All  &  0 & 0 & & 7 & 13 & & 0 & 4\vspace{-0.1cm}\\
\hline
\end{tabular}\\
\end{center}
$^ \mathrm{a}$Columns labelled as "Pure" represent the contribution of pure AGNs (Type 1 and 2) to each HSF galaxy type (i.e., without starburst connection).\\
$^\mathrm{b}$Columns labelled as "Total" provide the total contribution of AGNs to each type, including pure AGNs (Type 1 and 2) and mixed types (starburst-dominated AGNs, starburst-contaminated AGNs, and normal galaxies hosting AGNs). 
\end{minipage}
\end{table}

Active Galactic Nuclei (AGN) emission can coexist with star formation in galaxies, i.e., both phenomena are not mutually exclusive. Moreover, AGN activity is expected to be induced and enhanced in gas-rich mergers \citep{2008ApJS..175..390H,2008ApJS..175..356H}, so we should expect to find AGNs in our red galaxy sample, especially associated to our irregular class. Galaxies with an AGN are expected to be bright in the mid-to-far IR (and, in particular, in 24 $\mu m$) due to the emission of the hot dust surrounding the central engine \citep[see references in][]{2005ApJ...630...82P}. Therefore, pure or dominant AGN emitters can be confused with HSF galaxies according to the criterion used in this study to identify HSF galaxies. 

Therefore, we have cross-correlated our red galaxy sample with the AGN catalogue over the Extended Groth Strip by \citet{2009AJ....137..179R} to estimate the contamination of our HSF class by pure or dominant AGNs. These authors use a reliable technique of classification of active galactic nuclei (AGNs) based on fits of the UV-to-FIR SEDs of the galaxies with a complete set of AGN and starburst galaxy templates, in such a way that they are capable of distinguishing between pure and host-dominated AGNs up to $z\sim 2$ . The contribution of AGNs of any type and of pure AGNs (i.e., with no contribution of starbursts) to our HSF red galaxy sample at each redshift bin is tabulated in Table\,\ref{Tab:agns}. We also list the contribution of AGNs to HSF red galaxies at each redshift bin by morphological types. 

The results in Table\,\ref{Tab:agns} show that the contamination by any type of AGNs to our HSFs is negligible (below 13\% at all redshifts), the contribution of pure AGNs being even smaller ($<7$\% at all redshifts). Therefore, we can conclude that AGN contamination does not affect our results concerning the classification into HSF galaxies significantly. 

\section{Errors and uncertainties}
\label{Sec:Error}

The errors in the estimates of the number density of each red galaxy type are derived considering the following sources of errors: statistical and classification errors, photometric redshift errors, and cosmic variance uncertainties. 

Statistical counting errors have been estimated for the number densities derived by each classifier for each galaxy type and at each redshift \citep[][]{1986ApJ...303..336G}. The final number density of each morphological type at each redshift bin is estimated as the mean of the number densities resulting from the three independent classifiers. We have estimated the error of this mean 
as the quadratic propagation of the statistical error of each classifier \citep[consult][]{2009ApJ...697.1971J}. 

Concerning to redshift errors, we have used the redshift estimates from the Rainbow Extragalactic Database \citep{2011ApJS..193...30B}. As commented in \S\ref{Sec:sample}, uncertainties in the redshifts are $\Delta z / (1+z) \lesssim 0.03$ for the whole redshift interval under consideration, for both bright and faint red sources. 

We have included estimates of the uncertainties introduced in the number densities of each red galaxy type by the redshift errors. These estimates have been derived using Monte-Carlo simulations. We have made 100 simulated catalogues of the (mass-limited) red galaxy sample, adopting a photometric redshift value for each source between $[z_\mathrm{phot} - \Delta(z_\mathrm{phot}), z_\mathrm{phot} + \Delta(z_\mathrm{phot})]$ at random, being $z_\mathrm{phot}$ the nominal photometric redshift of the source and $\Delta(z_\mathrm{phot})$ the typical dispersion of the photometric redshifts compared to the spectroscopic ones. This dispersion is estimated as: $\Delta(z_\mathrm{phot}) = < \Delta(z)/(1+z)> \cdot (1+z_\mathrm{phot}$), where $<\Delta(z)/(1+z)>$ is the average value obtained for this normalized dispersion at the redshift bin of the galaxy (see Fig.\,\ref{Fig:zspec-zphot}). Then, we have obtained the number densities corresponding to each simulated catalogue for each galaxy type and at each redshift bin, accounting for the different redshifts of each catalogue. The dispersion of the 100 values obtained for the number density at each redshift bin and galaxy type represents an estimate of the error associated to the photometric redshift uncertainties. 

Statistics of massive red galaxies can be dramatically affected by cosmic variance due to their high clustering \citep{2004ApJ...600L.171S}. We have estimated cosmic variance using the model by \citet{2011ApJ...731..113M}, which provides estimates of cosmic variance for a given galaxy population using predictions from cold dark matter theory and the galaxy bias. They have developed a simple recipe to compute cosmic variance for a survey as a function of the angular dimensions of the field and its geometry, the mean redshift and the width of the considered redshift interval, and the stellar mass of the galaxy population. We have considered the geometry and angular dimensions of our field, as well as the different redshift bins analysed in each case to estimate the cosmic variance. \citeauthor{2011ApJ...731..113M} software provides these estimates in two mass ranges overlapping with ours: $10.5<\log (\mathcal{M}_*/\Msun)<11$ and $11<\log (\mathcal{M}_*/\Msun)<11.5$. Therefore, we have considered the mean cosmic variance of both mass ranges as a representative value of the cosmic variance of our mass-limited sample at each redshift bin. Cosmic variance depends on the redshift. At $0.3<z<1.5$ and for our mass range, cosmic variance decreases from $0.3<z<0.5$ to $0.7<z<0.9$ approximately, and then starts rising again towards higher redshifts. This behaviour is observed for all used redshift bins ($\Delta z = 0.1$, 0.2, and 0.4). For $\Delta z = 0.1$, the root cosmic variance acquires minimum and maximum values equal to 34\% and 40\%, respectively. For $\Delta z = 0.2$, it changes between 24\% and 28\%, and between 17\% and 19\% for $\Delta z=0.4$. All our results include these uncertainties quadratically added to the other error sources (statistical and classification errors, and redshift uncertainties).

The predictions of the cosmic variance obtained with the model by \citet{2011ApJ...731..113M} are in good agreement with the rough estimates that can be derived from panel a of Fig.\,\ref{Fig:comparacion}. At $0.7<z<1.1$, the dispersion of the different studies is $\sim 30$\%, most of which can be attributed to cosmic variance. This value is quite similar to the estimate obtained with \citeauthor{2011ApJ...731..113M} model at these redshifts for $\Delta z = 0.2$ ($\sim 27$\%). At lower and higher redshifts, the dispersion among different authors may arise also in  completeness problems and higher observational errors, so a direct comparison cannot be done.

The redshift errors have been added quadratically to the statistical and classification errors and to the uncertainties associated to cosmic variance to obtain the final error of the number density at each $z$ interval and for each galaxy type. In general, for regular red galaxies, the statistical and classification errors contribute to $\sim 30$\% to the total error of their number density for both low and high redshifts, the redshift errors represent $\sim 10$\% of the total error, and the cosmic variance contributes to $\sim 60$\% of the total error at all redshifts. For irregular galaxies, the contribution of the statistical and morphological classification errors represents $\sim 35$\% at all redshifts, the redshift errors are $\sim 25$\% (being $\sim 30$\% at low z, and decreasing down to $\sim 23$\% at high z), and the cosmic variance contributes to $\sim 37$\% (being $\sim 35$\% at low z, and rising to $\sim 40$\% at high z).

\begin{table*}
\begin{minipage}{\textwidth}
\caption{Correspondence between different galaxy evolutionary stages and our red galaxy types.\label{Tab:correspondence}}
\begin{center}
\begin{tabular}{llll}
\hline
\multicolumn{1}{c}{Galaxy Type} & \multicolumn{1}{c}{Global morphology} & \multicolumn{1}{c}{Structural distortion$^\mathrm{a}$} & \multicolumn{1}{c}{Star formation$^\mathrm{b}$} \\
\multicolumn{1}{c}{(1)} & \multicolumn{1}{c}{(2)} &\multicolumn{1}{c}{(3)} &\multicolumn{1}{c}{(4)} \\\hline 
E-S0 & Spheroid-dominated &  Regular & LSF \\
Spiral & Disk-dominated       &  Regular   & HSF/LSF \\
\hline\vspace{-0.2cm}\\
Gas-rich major mergers (merging-nuclei phase)& Disk-dominated      & Irregular  & HSF/LSF \\
Gas-rich major mergers (post-merger phase)   & Spheroid-dominated  & Irregular  & HSF/LSF \\
Gas-poor major mergers & Spheroid-dominated  & Irregular  & LSF \\\hline
\end{tabular}\\
\end{center}
$^ \mathrm{a}$Non-distorted morphologies are noted as regular, and merger-like distorted ones as irregulars.\\
$^ \mathrm{b}$HSF: objects with enhanced star formation compared to the average of the whole galaxy population; LSF: objects with star formation lower than this average SFR.\\
$^ \mathrm{c}$Only major mergers already merged into one body are traced by this study. Galaxy pairs are considered as two still-independent galaxies.\\
\end{minipage}
\end{table*}

\section{Observational tests to the hierarchical origin of massive E-S0's}
\label{Sec:Tests}

As commented in \S\ref{Sec:introduction}, no observational evidence on the existence of a evolutionary link between major mergers and the rise of the present-day massive E-S0's exists up to the date. In order to test the major-merger origin of massive E-S0's observationally, we must be capable of providing evidence supporting or rejecting the existence of such a link. However, the transitory stages and end products of these evolutionary tracks coexist at each redshift, making this task difficult. Here we define three tests to observational data based on the predictions of hierarchical models to check if data are coherent with the existence of this link, that can be done thanks to the classification performed in this study.

\subsection{Equivalence between our galaxy types and the different evolutionary stages of red galaxies}
\label{Sec:Equivalence}

The morphological and structural properties of a galaxy can be combined with information about its star formation enhancement to establish correspondences between its galaxy type in our classification and the evolutionary stage of the galaxy. Once a galaxy is identified as a major merger on the basis of its noticeable structural distortion, both its star formation level and its global morphology may be pointing either to the gas-content of its progenitors or to the merger phase \citep{2008MNRAS.384..386C,2008MNRAS.391.1137L,2010MNRAS.404..590L,2010MNRAS.404..575L}. 

Galaxies involved in a gas-rich major merger are expected to be disk-dominated during the merging-nuclei phase. During the initial moments of this phase, these mergers experience strong starbursts that last less than $\sim 0.5$\,Gyr. Therefore, depending on the evolutionary stage of these mergers, they might appear either as HSF or LSF. Post-merger stages of these events will still exhibit noticeable distortion, but its morphology is expected to be spheroidal-dominated. Depending on the efficiency of the star formation quenching in the merger, late phases of gas-rich major mergers may be HSF or LSF. Additionally, gas-poor mergers are expected to be irregular, spheroidal-dominated, and mostly LSF since their merging-nuclei phase. Normal disks and E-S0's exhibit a regular disk or spheroidal-dominated morphology, respectively. Most E-S0's are quiescent, meaning that they must be LSF.

Our classification thus allows us to distinguish among these evolutionary stages of red galaxies just attending to their types, except between gas-poor mergers and post-merger stages of wet mergers. In Table\,\ref{Tab:correspondence} we list the correspondence between a galaxy evolutionary stage and its characteristics in our classification.

\subsection{Hierarchical evolutionary paths among red galaxy types}
\label{Sec:F07}

Here, we describe three tests based on the expectations of hierarchical models of galaxy formation to check whether data is coherent with the existence of an evolutionary link between major mergers and the E-S0's appearing on the Red Sequence since $z\sim 1.5$ or not.

F07 proposed a hierarchical mixed evolutionary scenario for explaining the observed mass migration from the massive end of the Blue Galaxy Cloud to that of the Red Sequence since $z\sim 2$, in which "quenched galaxies enter the Red Sequence via gas-rich mergers", and can be "followed by a limited number of gas-poor, stellar mergers along the sequence". The semi-analytical model by \citet[EM10 hereafter]{2010A&A...519A..55E} proved the feasibility of the F07 scenario for explaining the buildup of E-S0's that end up with $\mathrm{M}_*>10^{11}\Msun$ at $z=0$, just accounting for the effects of the major mergers strictly reported by observations since $z\sim 1.2$ \citep{2009ApJ...694..643L}. This model reproduces the observed evolution of the massive end of the galaxy luminosity function by color and morphological types. The evolutionary track described by F07 appears naturally in the model, as it considers the relative contribution of gas-poor and gas-rich mergers at each redshift reported by \citet{2008ApJ...681..232L} and their different effects on galaxy evolution. 

The advantage of this model is that its predictions are in excellent agreement with cosmological hierarchical models (despite being based on observational major merger fractions), reproducing observational data at the same time \citep[see EM10;][]{2010arXiv1003.0686E}. Based on these predictions, we have defined some tests that observational data must fulfill if most massive E-S0's have really derived from major mergers occurred at relatively late epochs in the cosmic history. These predictions are the following ones:

\begin{enumerate}
 \item Most present-day E-S0's with $\mathrm{M}_*>10^{11}\Msun$ are the result of at least one gas-rich major merger that place them on the Red Sequence since $z\sim 1.2$.
 \item In addition, $\sim 75$\% of the remnants resulting from these gas-rich events have been involved in a subsequent gas-poor major merger, occurred quite immediately. The remaining $\sim 25$\% have thus continued their evolution towards an E-S0 passing through a quiet post-merger phase. 
 \item The bulk of these major mergers are at intermediate-to-late stages during the $\sim 2$\,Gyr period elapsed at $0.7<z<1.2$, which means that the gas-rich ones must have started at $1<z<1.5$, accouting for the typical timescales of these events. The gas-poor ones are later, but must take place earlier than $z\sim 0.7$ in any case, as the resulting massive E-S0's have been in place since that epoch. 
\end{enumerate}

Therefore, according to the EM10 model, the appearance of the bulk of massive E-S0's takes place at $0.7<z<1.2$, and nearly all have evolved according to the following path: 

\begin{center}
 {\small
 \framebox[0.26\textwidth][c]{
  \begin{minipage}[c]{0.25\textwidth}
  \begin{center}
 {\bf Gas-rich major merger\/} \\  
  \end{center}
  \end{minipage}
 } 

\vspace{0.2cm}
$\Downarrow$
\vspace{0.2cm}

\begin{tabular}{lr}

\framebox[0.22\textwidth][c]{
  \begin{minipage}[c]{0.21\textwidth}
   \begin{center}
{\bf Post-merger stage\/}\\
$[$ $\sim 25$\% $]$\\
  \end{center}
  \end{minipage}
 } 
&

\framebox[0.22\textwidth][c]{
  \begin{minipage}[c]{0.21\textwidth}
   \begin{center}
{\bf Gas-poor major merger\/}\\
$[$ $\sim 75$\% $]$\\
  \end{center}
  \end{minipage}
 } 

 \\
\end{tabular}

 \vspace{0.2cm}
$\Downarrow$
\vspace{0.2cm}

\framebox[0.2\textwidth][c]{
  \begin{minipage}[c]{0.18\textwidth}
   \begin{center}
{\bf Massive E-S0a\/} 
  \end{center}
  \end{minipage}
 }. \vspace{-0.5cm}
\begin{equation}\label{eq:track}
\end{equation}
} 
\end{center}
 
\noindent Note that this evolutionary track implies the existence of a nearly 1:1:1 numerical relation at $0.7<z<1.2$ between gas-rich major mergers at merging-nuclei stages, gas-poor events and post-mergers stages of gas-rich ones, and the massive E-S0's assembled at those epochs. Accounting for the correspondence of these evolutionary stages and the red galaxy types defined in this study (see Table\,\ref{Tab:correspondence}), the evolutionary path schematized in eq.\,\ref{eq:track} can be re-written as follows:

\begin{center}
 {\small
 \framebox[0.12\textwidth][c]{
  \begin{minipage}[c]{0.11\textwidth}
  \begin{center}
 {\bf ID (HSF/LSF)\/} \\  
  \end{center}
  \end{minipage}
 } 

\vspace{0.2cm}
$\Downarrow$
\vspace{0.2cm}

\begin{tabular}{lr}

\framebox[0.15\textwidth][c]{
  \begin{minipage}[c]{0.14\textwidth}
   \begin{center}
{\bf IS (HSF/LSF)\/}\\
$[$ $\sim 25$\% $]$\\
  \end{center}
  \end{minipage}
 } 
&

\framebox[0.15\textwidth][c]{
  \begin{minipage}[c]{0.3\textwidth}
   \begin{center}
{\bf IS (mostly LSF)\/}\\
$[$ $\sim 75$\% $]$\\
  \end{center}
  \end{minipage}
 } 

 \\
\end{tabular}

 \vspace{0.2cm}
$\Downarrow$
\vspace{0.2cm}

\framebox[0.18\textwidth][c]{
  \begin{minipage}[c]{0.17\textwidth}
   \begin{center}
{\bf RS (mostly LSF)\/} 
  \end{center}
  \end{minipage}
 }. \vspace{-0.5cm}
\begin{equation}\label{eq:track2}
\end{equation}
} 
\end{center}

\noindent Then, the previous 1:1:1 relation between the different stages of these galaxies in their evolution towards the Red Sequence can be translated into an equivalent 1:1:1 relation between the following red galaxy types, according to Table\,\ref{Tab:correspondence}:

\begin{equation}\label{eq:track3}
 \mathrm{\textbf{ID}} \rightarrow \mathrm{\textbf{IS}} \rightarrow \mathrm{\textbf{RS}}.  
\end{equation}

\noindent If most massive E-S0's have really evolved according to the hierarchical scenario proposed by the EM10 model, then the data must fulfill this evolutionary path among massive red galaxy types at $0.7<z<1.2$. This imposes the following three observational tests or constraints to observations:

\begin{itemize}
\item[1.] The accumulated number density of ID's (gas-rich major mergers) on our red sample since $z\sim 1.2$ down to a lower redshift $z$ must reproduce the net numerical increment of RS's (E-S0's) observed between the two redshifts. 

\item[2.] An analogous relation must be fulfilled by the accumulated number density of IS's (gas-poor major mergers and post-merger stages of gas-rich events).

\item[3.] The bulk of RS's (E-S0's) with stellar masses $\mathrm{M}_*>10^{11}\Msun$ at $z=0$ must have been definitely built up during the $\sim 2.2$\,Gyr time period elapsed at $0.7<z<1.2$.
\end{itemize}

Our data is complete for galaxy masses $\mathrm{M}_*>5\times10^{10}\Msun$ at each redshift. This means that we can ensure that we are in a position to trace back in time the potential progenitors of the present-day E-S0's with $\mathrm{M}_*>10^{11}\Msun$ at $z\sim 0$ that could have merged to create them during the last $\sim 9$\,Gyr (see \S\ref{Sec:sample}).

Therefore, we have estimated the cumulative distribution of ID's and of IS's, and we have compared them with the redshift evolution of the number density of RS's since $z\sim 1.5$. In the case that major mergers have not driven the assembly of the massive E-S0's as proposed by the EM10 model, the previous three tests must fail. On the contrary, if these three distributions agree pretty well, these tests will support strongly the existence of an evolutionary link between major mergers and the appearance of massive E-S0's, as expected by hierarchical scenarios of galaxy formation. The results of these tests are presented in \S\ref{Sec:TestsResults}. 

We must remark that the EM10 model exclusively quantifies the effects of major mergers on galaxy evolution at $z<1.2$. Hence, it does not discard the contribution of different evolutionary processes to the definitive assembly of massive E-S0's, although it predicts that it must have been low, as most of their number density evolution can be explained just accounting for the effects of the major mergers. This seems to be confirmed by observations, as other evolutionary mechanisms (such as minor mergers, ram pressure stripping, or bars) seem to have been significant for the formation of the Red Sequence only for low and intermediate masses, but not for high masses \citep[][]{2004ARA&A..42..603K,2007ApJ...660.1151D,2008A&A...489.1003D,2009ApJ...697.1369B,2009A&A...508.1141S,2010MNRAS.409..346C,2011MNRAS.411.2148K}. Moreover, the model does not exclude disk rebuilding after the major merger either. On the contrary, it is probably required for giving rise to a S0 instead of an elliptical, as indicated by observations \citep{2005A&A...430..115H,2009A&A...507.1313H,2009A&A...496..381H,2009A&A...501..437Y}.

Moreover, the EM10 model assumes that intermediate-to-late stages of major mergers are red and will produce an E-S0, on the basis of many observational and computational studies \citep[see references in EM10 and][]{2002A&A...381L..68C,2007MNRAS.382.1415S,2010ApJ...714L.108S,2010A&A...518A..61C}. These assumptions are crucial for the model, as they are necessary to reproduce the redshift evolution of the luminosity functions selected by color and morphological type. Therefore, testing if our data is coherent with the existence of an evolutionary link between the advanced stages of major mergers in our red sample and the definitive buildup of massive E-S0's, we are also indirectly testing these assumptions of the EM10 model.

\subsection{Observational considerations for the tests}
\label{Sec:RSEvolution}

According to \citet{2006ApJ...652..270B}, the average number density of major mergers at a given redshift centered at $z$, $n _\mathrm{m}(z)$, is related to the number density of the major mergers detected at certain intermediate phase of the encounter in that redshift bin, $<n_\mathrm{m,phase}(z)>$, as follows:

\begin{equation}\label{eq:density}
n_\mathrm{m}(z) = <n_\mathrm{m,phase}(z)>  \frac{t([z_1,z_2])}{\tau _\mathrm{det}},
\end{equation}

\noindent with $t([z_1,z_2])$ being the time elapsed in the redshift bin, and $\tau _\mathrm{det}$ representing the  detectability time of the intermediate merger stage under consideration. We have used this equation in our tests.

We find that the number densities of ID's and IS's remain quite constant with redshift (see \S\ref{Sec:Results}). As ID's and IS's correspond to intermediate merger stages, this means that the major merger rate must evolve smoothly with redshift, in good agreement with observational estimates of merger rates  \citep{2008ApJ...681..232L,2011ApJ...739...24B,2011ApJ...742..103L}. This also indicates that the net flux of irregular galaxies appearing on the Red Sequence or at nearby locations on the Green Valley (i.e., the number of red irregulars created and destroyed per unit time in the red sample) must have been nearly constant at $0.3<z<1.5$ for our mass range. Together with the evolutionary track stated in eq.~\ref{eq:track3}, this implies that the accumulated number of ID's observed in a redshift bin must be equal to the accumulated number of IS's in the same bin, and also to the number of RS's assembled at the end of the redshift bin. As we also find that the number of RS's is negligible at $z=1.5$ (see \S\ref{Sec:Results}), it is also justified to start accumulating ID's and IS's since this redshift down to $z\sim 0.3$.

In order to estimate $n_\mathrm{m}(z)$, we have adopted the merger timescales derived by \citet{2010MNRAS.404..590L}. These authors report typical detectability timescales through CAS indices of simulated major mergers, as a function of the baryonic gas fraction and the mass ratio of the encounter. For gas fractions representative of gas-rich mergers up to $z\sim 1$ \citep[$\sim 30-50\%$, see][]{2009A&A...507.1313H} and mass ratios typical of major mergers (1:1 to 1:4), they find an average detectability timescale $\tau _\mathrm{det,gas-rich}\sim 1.0\pm 0.3$\,Gyr of merging-nuclei phases. For gas fractions representative of gas-poor major mergers ($\lesssim 30$\%), we estimated from these author and from \citet{2005AJ....130.2647V} an average timescale of $\tau _\mathrm{det,gas-poor}\sim 0.5\pm 0.2$\,Gyr of their merging-nuclei phase. The timescales for post-merger stages are similar to those of their merging-nuclei phases for both gas-rich  and gas-poor mergers
 \citep{2007AJ....133.2132S,2010MNRAS.404..590L}. The errors in these timescales account for small changes of these representative values with redshift, accordingly to \citet{2010ApJ...718.1158L}. 

\begin{table*}
\begin{minipage}{\textwidth}
\caption{Comoving number densities of the different red galaxy types in units of $\times 10^{-4}$ Mpc$^{-3}$.\label{Tab:densities}}
\begin{center}
\begin{tabular}{lcrrcrrcrrcrrcr}
\hline
\multicolumn{1}{c}{\multirow{2}{*}{Redshift}} & \multicolumn{1}{c}{\multirow{2}{*}{Number$^ \mathrm{a}$}} &\multicolumn{2}{c}{N(RS)$^\mathrm{b}$} && \multicolumn{2}{c}{N(RD)}  &&  \multicolumn{2}{c}{N(IS)} && \multicolumn{2}{c}{N(ID)}  && \multicolumn{1}{c}{\multirow{2}{*}{N(Total)}}  \vspace{0.2cm}\\ \cline{3-4}\cline{6-7}\cline{9-10}\cline{12-13}
\multicolumn{1}{c}{} & \multicolumn{1}{c}{} &\multicolumn{1}{c}{LSF$^\mathrm{c}$} & \multicolumn{1}{c}{HSF} && \multicolumn{1}{c}{LSF} & \multicolumn{1}{c}{HSF} &&  \multicolumn{1}{c}{LSF} & \multicolumn{1}{c}{HSF} && \multicolumn{1}{c}{LSF} & \multicolumn{1}{c}{HSF}  && \multicolumn{1}{c}{}  \\
\multicolumn{1}{c}{(1)} & \multicolumn{1}{c}{(2)} & \multicolumn{1}{c}{(3)} & \multicolumn{1}{c}{(4)} &&   \multicolumn{1}{c}{(5)}& \multicolumn{1}{c}{(6)} && \multicolumn{1}{c}{(7)} & \multicolumn{1}{c}{(8)} && \multicolumn{1}{c}{(9)} & \multicolumn{1}{c}{(10)}  && \multicolumn{1}{c}{(11)}\\\hline
$0.3<z<0.5$ & 18& 5.2$^{+2.1}_{-1.8}$ & 1.2$^{+1.0}_{-0.6}$ && 0.6$^{+0.9}_{-0.9}$ & 2.3$^{+1.3}_{-0.9}$ && 0.0 & 0.8$^{+1.5}_{-1.5}$ && 0.0 & 1.9$^{+1.2}_{-0.8}$ &&  11.0$^{+3.4}_{-2.7}$ \\[0.2cm]
$0.5<z<0.7$ & 48&10.2$^{+2.9}_{-2.8}$ & 1.7$^{+0.8}_{-0.6}$ && 1.7$^{+0.8}_{-0.6}$ & 1.2$^{+0.7}_{-0.5}$ && 0.8$^{+0.7}_{-0.5}$ & 0.2$^{+0.8}_{-0.8}$ && 1.0$^{+0.6}_{-0.4}$ & 0.1$^{+0.4}_{-0.4}$ && 16.6$^{+3.4}_{-3.2}$ \\[0.2cm]
$0.7<z<0.9$ & 61& 6.5$^{+1.8}_{-1.7}$ & 1.6$^{+0.9}_{-0.8}$ && 1.6$^{+0.7}_{-0.5}$ & 0.3$^{+0.3}_{-0.3}$  &&0.9$^{+0.6}_{-0.6}$ & 1.6$^{+1.1}_{-0.9}$ && 0.3$^{+0.3}_{-0.3}$	  & 2.8$^{+0.9}_{-0.8}$ && 15.0$^{+2.7}_{-2.4}$\\[0.2cm]
$0.9<z<1.1$ &54 & 3.4$^{+1.0}_{-1.0}$ & 0.3$^{+0.3}_{-0.2}$ && 0.3$^{+0.3}_{-0.2}$ & 0.2$^{+0.3}_{-0.3}$	 &&2.2$^{+0.8}_{-0.7}$ & 0.9$^{+0.7}_{-0.5}$ && 0.9$^{+0.4}_{-0.3}$	  & 2.3$^{+0.8}_{-0.7}$ && 10.6$^{+1.8}_{-1.6}$\\[0.2cm]
$1.1<z<1.3$ & 31&1.0 $^{+0.4}_{-0.3}$ & 0.2$^{+0.2}_{-0.1}$ && 0.1$^{+0.2}_{-0.2}$ & 0.0  &&1.2$^{+0.5}_{-0.4}$ & 0.9$^{+0.7}_{-0.5}$ && 0.8$^{+0.4}_{-0.3}$	  & 1.1$^{+0.5}_{-0.4}$ && 5.3$^{+1.2}_{-0.9}$\\[0.2cm]
$1.3<z<1.5$ & 24&0.2$^{+0.2}_{-0.1}$ & 0.0 && 0.0 & 0.0  &&1.0$^{+0.5}_{-0.5}$ & 0.7$^{+0.6}_{-0.5}$ && 0.4$^{+0.3}_{-0.2}$	  & 1.5$^{+0.6}_{-0.5}$ && 3.7$^{+1.1}_{-0.9}$\\
\hline
\end{tabular}\\
\end{center}
$^\mathrm{a}$Net number of galaxies in the final sample in a sky area of $\sim 155$\,arcmin$^{2}$.\\
$^\mathrm{b}$Galaxy types according to morphology: Regular Spheroid-dominated (RS), Regular Disk-dominated (RD), Irregular Spheroid-dominated (IS), Irregular Disk-dominated (ID).\\
$^\mathrm{c}$Galaxy types according to star formation activity: High Star-Forming (HSF), Low Star-Forming (LSF).
\end{minipage}
\end{table*}

\begin{figure*}
\begin{center}
\includegraphics[width=\textwidth,angle=0]{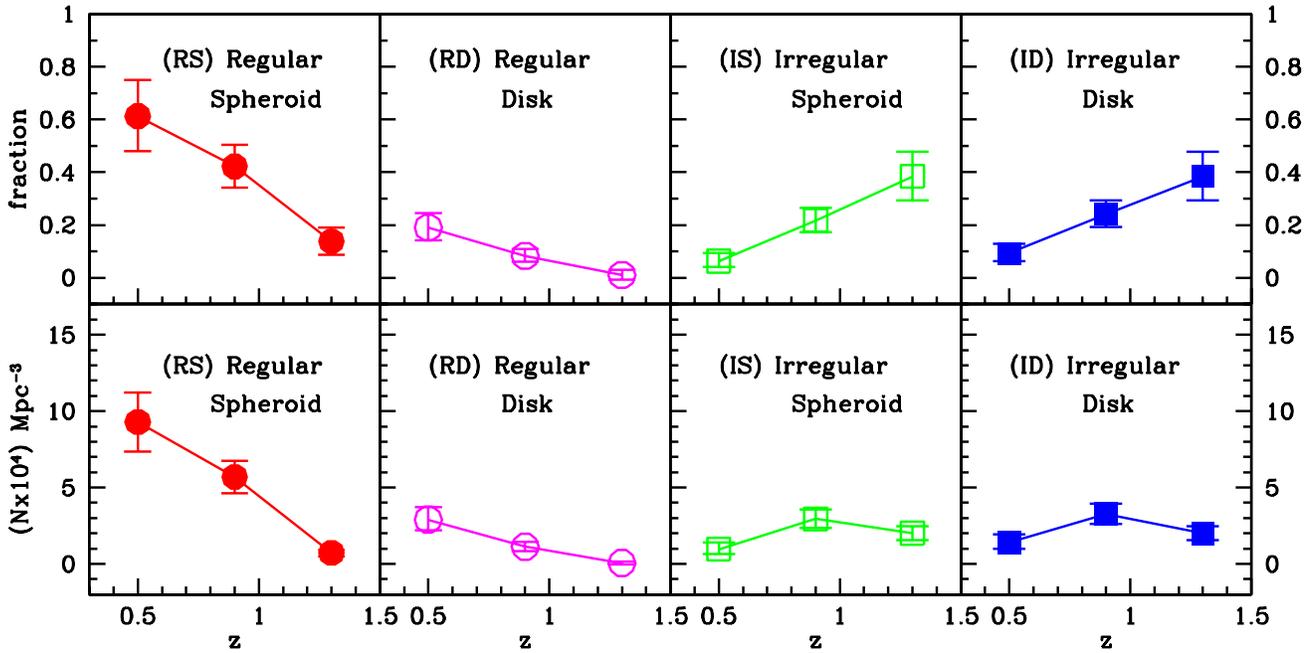} 
\caption{Redshift evolution of the number density and fraction of red galaxies with $\mathrm{M}_*>5\times 10^{10}\Msun$, according to their morphology and structural distortion level, for the three wide redshift bins considered in the study ($0.3<z<0.7$, $0.7<z<1.1$, and $1.1<z<1.5$). \emph{Top panels}: Redshift evolution of the fraction of each type with respect to the whole red galaxy population at each redshift bin. \emph{Bottom panels}: Redshift evolution of the number density of each morphological type in the same redshift bins. 
}\label{Fig:fractypes}
\end{center}
\end{figure*}

\section{Results}
\label{Sec:Results}

In Table\,\ref{Tab:densities} we show the number densities of red galaxies with $\mathrm{M}_*>5\times 10^{10}\Msun$ per redshift bins at $0.3<z<1.5$ obtained from our data, according to the classification resulting from the morphological/distortion level of the galaxy (\S\ref{Sec:Morphology}) and from its star formation level (\S\ref{Sec:ClassificationSED}). Therefore, the galaxies are classified as regulars or irregulars (according to their structural distortion level), as spheroid/disk-dominated (according to their global morphology), and as HSF/LSF (according to their star formation enhancement). Errors in the results listed in this table and in all the figures of the present section account for the statistical and classification errors, the redshift errors, and the cosmic variance uncertainties, as described in \S\ref{Sec:Error}. 

\begin{figure}
\begin{center}
\includegraphics[width=0.5\textwidth,angle=0]{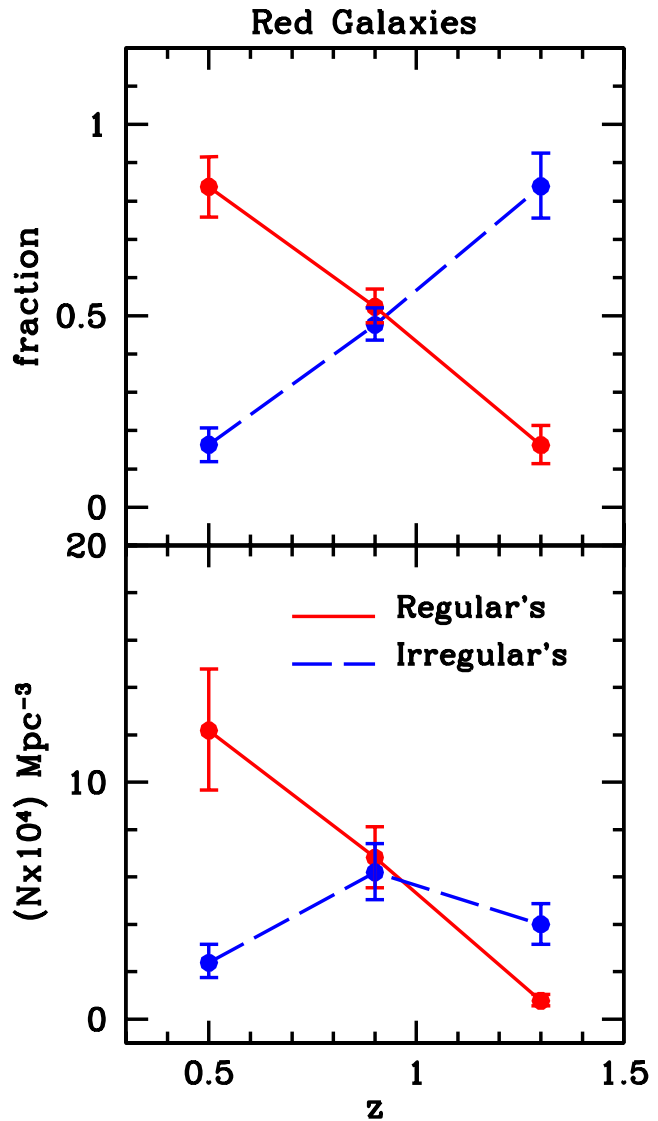}
\caption{Redshift evolution of the number density and fraction of all regular and irregular (i.e., major mergers) red galaxies with $\mathrm{M}_*>5\times 10^{10}\Msun$ at $0.3<z<1.5$, in the three wide redshift bins under consideration ($0.3<z<0.7$, $0.7<z<1.1$, and $1.1<z<1.5$). \emph{Top panel}: Redshift evolution of the fraction of red massive regular and irregular galaxies with respect to the total red galaxy population. \emph{Bottom panel}: Redshift evolution of the number density of regular and irregular massive red galaxies in the same redshift bins.} \label{Fig:E_D}
\end{center}
\end{figure}

\subsection{Morphological evolution of red galaxies since $z\sim 1.5$}
\label{Sec:MorphologyEvolution}

Figure\,\ref{Fig:fractypes} shows the redshift evolution of the fraction and the comoving number density of massive red galaxies by galaxy types considering their global morphology and structural distortion level. The fractions and number densities of red regular galaxies rise with cosmic time, with the fraction of spheroids being higher than that of disks at the same redshift. Therefore, we are tracing the progressive settlement of regular galaxies on the massive end of the Red Sequence at $0.3< z<1.5$, made primarily of spheroids but also containing some regular disks (mainly at $z<0.7$). We also find significant populations of irregular galaxies at all redshifts up to $z\sim 1.5$, with number densities $> 10^ {-4}$\,Mpc$^{-3}$. Although the fractions of red irregular spheroids and disks decrease with cosmic time, their densities remain quite constant at $0.3<z<1.5$. This fact clearly points to the transitory nature of red irregular galaxies at any redshift, as indicated by previous studies \citep{2008ApJ...672..177L}. 

Figure\,\ref{Fig:E_D} shows the redshift evolution of the number density and fractions of red regular galaxies (RS $+$ RD) and irregulars (IS $+$ ID) at $0.3<z<1.5$, for the three wide redshift bins under consideration. The fraction of regular galaxies increases by a factor of $\sim 6$ since $z\sim 1.3$ down to $z\sim 0.5$, whereas the fraction of irregulars decreases  by the same amount (top panel in the figure). Moreover, the number density of regulars has risen by a factor of $\sim 12$ during this time period (bottom panel in the figure), while that of the irregulars keeps constant until $z\sim 0.9$, decreasing by a factor of $\sim 3$ at $z\lesssim0.6$ \citep[in excellent agreement with the results by][]{2010ApJ...709..644I}.

At $z\sim 1.3$, irregular galaxies (i.e., major mergers) represent nearly 80\% of the red galaxy population. Their fraction decreases down to $\sim 50$\% at $z\sim 0.9$, and reaches $\lesssim 15$\% at $z\sim 0.5$. Considering that $\sim 70$\% of these irregular galaxies are dust-reddened due to intense star formation (see \S\ref{Sec:SFandMorphology}), these results are in good agreement with previous estimates of the fraction of dust-reddened star-forming galaxies on the Red Sequence at different redshifts \citep{2002A&A...381L..68C,2003ApJ...586..765Y,2003MNRAS.346.1125G,2004ApJ...600L..11B,2004ApJ...600L.131M,2005ApJ...620..595W,2007MNRAS.380..585C,2007A&A...465..711F,2011ApJ...726..110Z}.

\begin{figure}
\includegraphics[width=0.5\textwidth,angle=0]{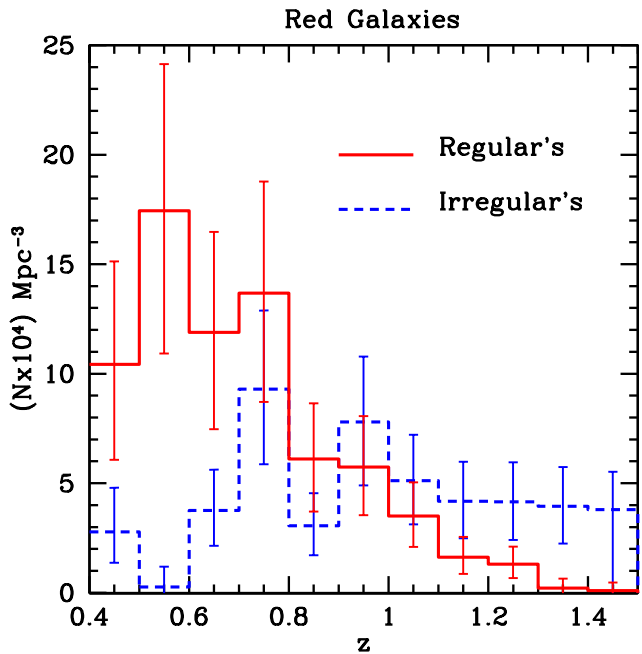}
\caption{Redshift evolution of the number density of red regulars and irregulars with $\mathrm{M}_*>5\times 10^{10}\Msun$ at $0.3<z<1.5$, using narrow redshift bins of $\Delta z = 0.1$.} \label{Fig:hist}
\end{figure}

We have derived the redshift evolution of the number density of regular and irregular massive red galaxies in narrower redshift bins, to delimit more accurately the epoch at which the bulk of the red regular galaxies appears into the cosmic scenario (see Fig.\,\ref{Fig:hist}). The large uncertainties of our results in these narrow redshift bins prevent us of deriving quantitative conclusions based on this figure. However, we can conclude from it that the number density of red irregulars (i.e., major mergers) has remained nearly constant at $0.7\lesssim z\lesssim 1.5$ within errors. This indicates that these systems must have been forming and disappearing at similar rates during this time period. Therefore, the decrease of the fraction of irregular systems with cosmic time is exclusively relative, i.e., it is just due to the rise of the population of regular galaxies, but not to a net drop in their number density down to $z\sim 0.6$. The figure also shows that the redshift-evolution of the number densities of regulars and irregulars on the Red Sequence cross at $z=0.9$ for $\mathrm{M}_*>5\times 10^{10}\Msun$. 

As commented in \S\ref{Sec:introduction}, the rise of the number density of red regular types with cosmic time and the constancy of that of irregular ones has been interpreted as a sign pointing to the conversion of irregulars into regulars with time. We provide observational evidence supporting the existence of this evolutionary link in \S\ref{Sec:TestsResults}. 

\begin{figure*}
\begin{center}
\includegraphics*[width=0.24\textwidth,angle=0]{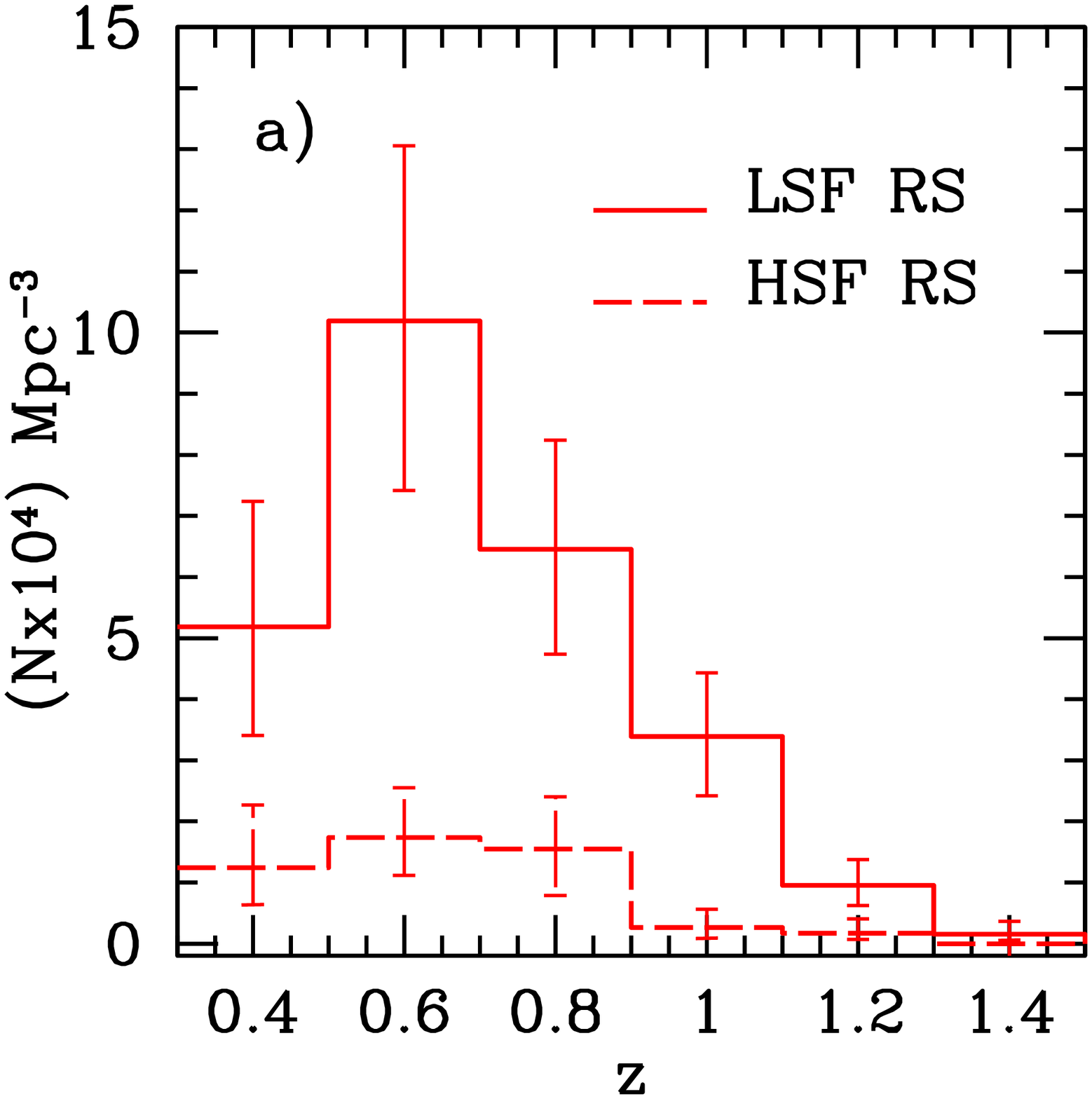}
\includegraphics*[width=0.24\textwidth,angle=0]{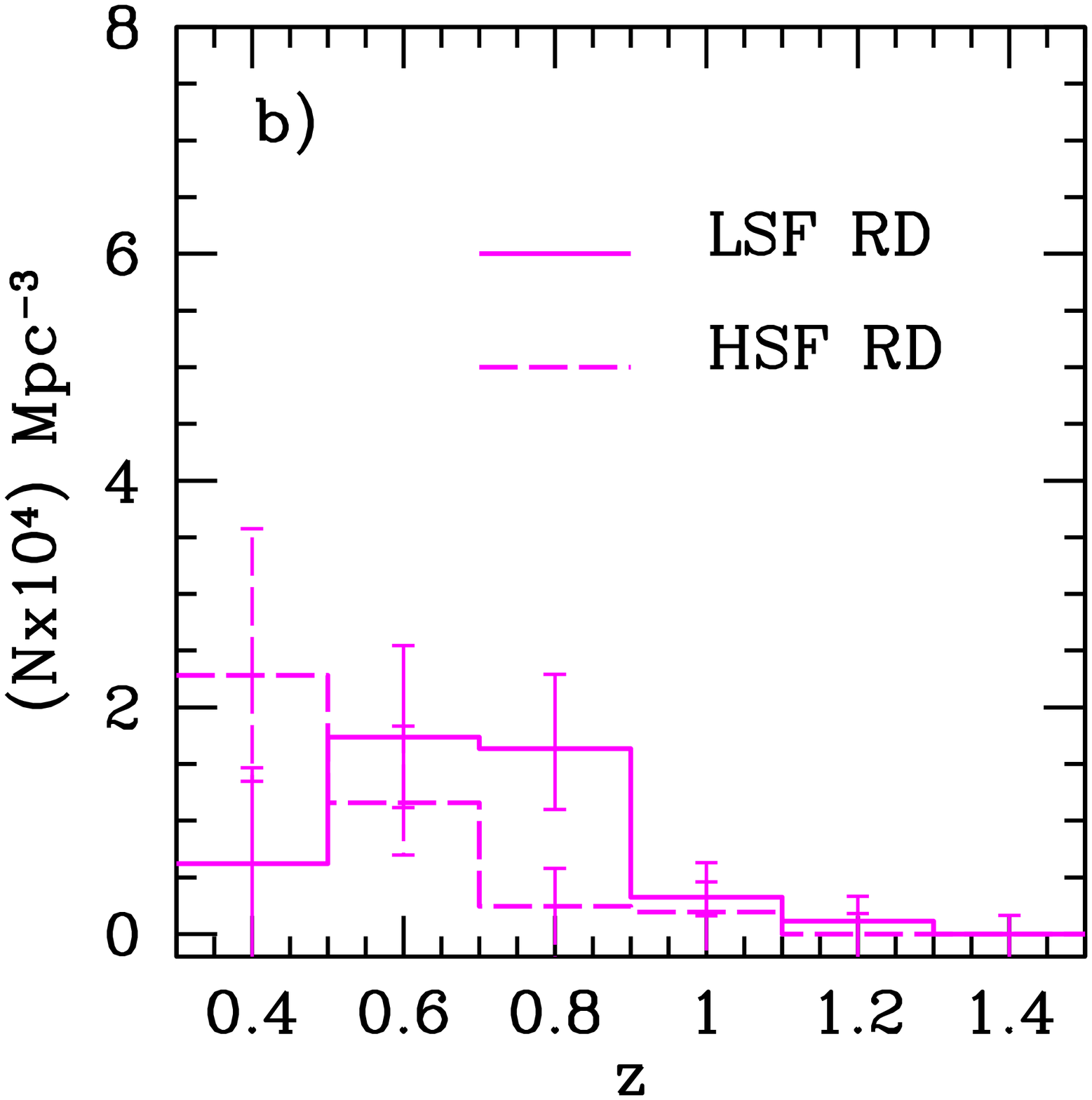}
\includegraphics*[width=0.24\textwidth,angle=0]{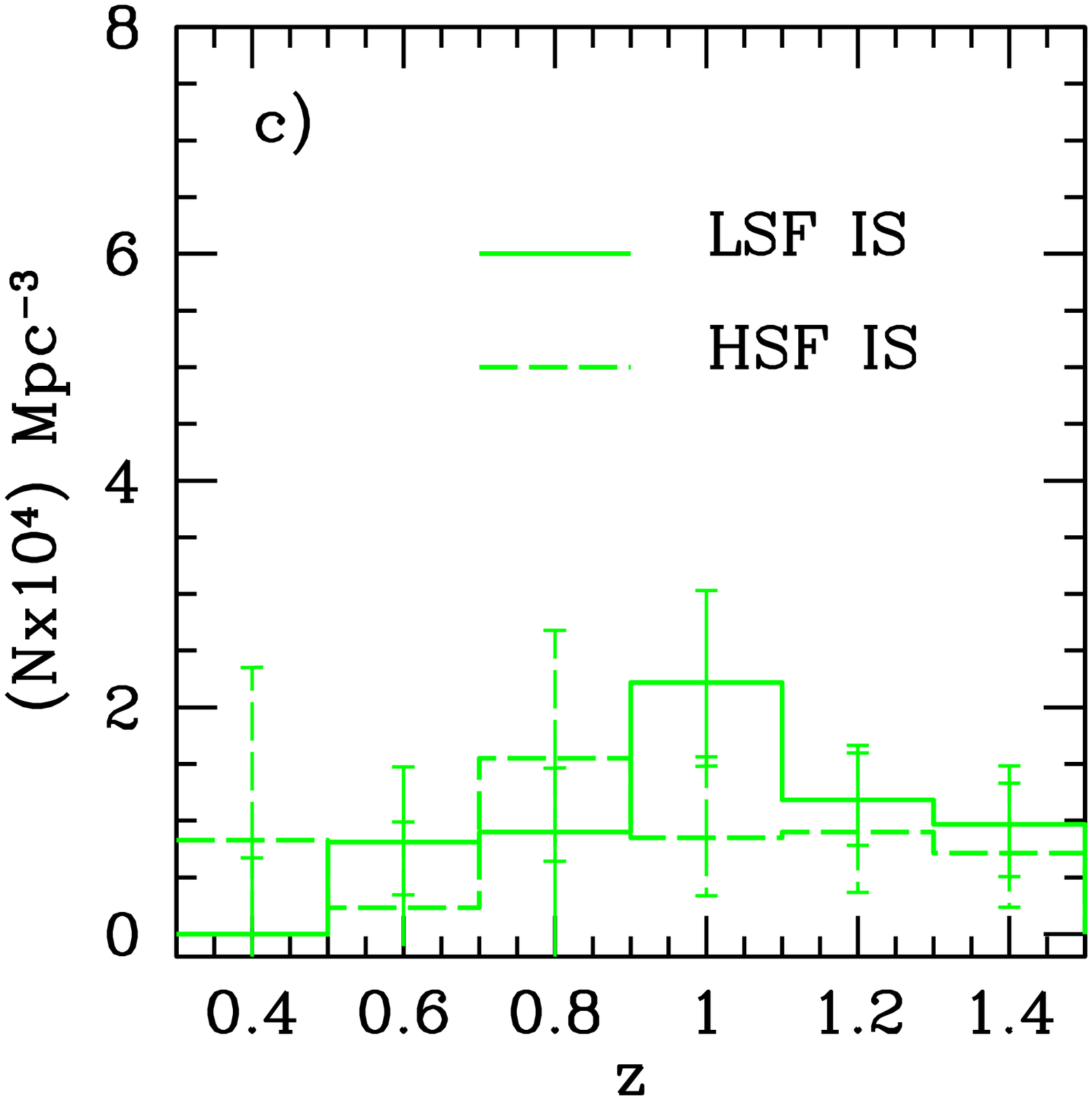}
\includegraphics*[width=0.24\textwidth,angle=0]{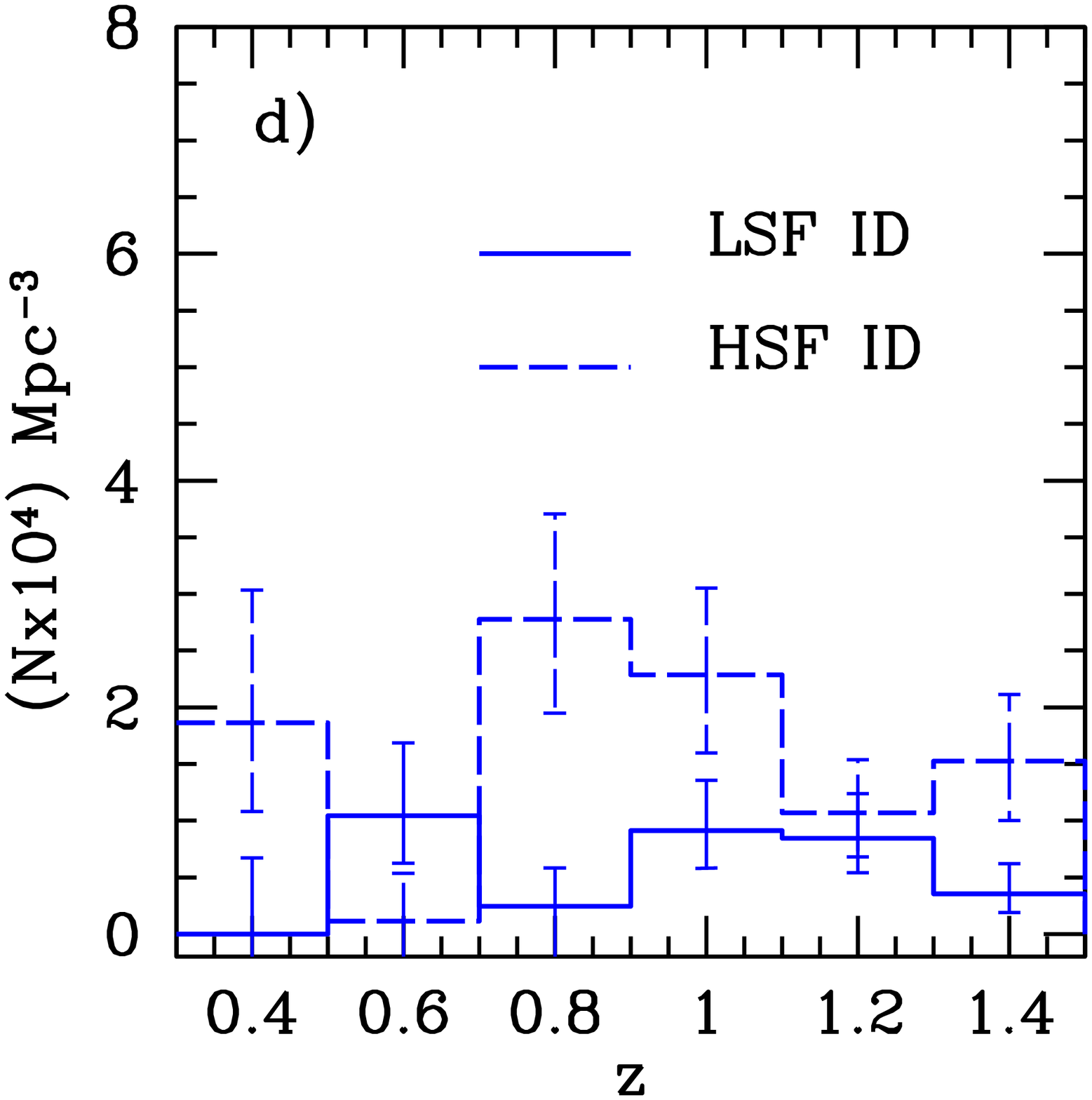}
\caption{Redshift evolution of the comoving number density of red galaxies with $\mathrm{M}_*>5\times 10^{10}\Msun$, according to their morphological, structural, and star formation activity properties. \emph{Solid lines}: Number densities for LSF galaxies of each type. \emph{Dashed lines}: Number densities of HSF galaxies of each morphological type. \emph{Panel a)}: Regular spheroids (RS's). \emph{Panel b)}: Regular disks (RD's). \emph{Panel c)}: Irregular spheroids (IS's). \emph{Panel d)}: Irregular disks (ID's). }\label{Fig:mip_p_classification}
\end{center}
\end{figure*}

\subsection{Star formation activity according to red galaxy types since $z\sim 1.5$}
\label{Sec:SFandMorphology}

In Fig.\,\ref{Fig:mip_p_classification} we show the redshift evolution  of the number density of massive red galaxies for each morphological type defined in this study, attending to its star formation activity. We remark that HSF's are galaxies that show enhanced SFRs compared to the average SFR of the galaxy population at each redshift (\S\ref{Sec:ClassificationSED}). All types, except RS's, host a significant number of HSF galaxies, with percentages varying depending on type and redshift. Red regular spheroids are the galaxy types hosting the lowest fractions of HSF systems at all redshifts since $z\sim 1.5$, as expected. Curiously, red RD's exhibit a noticeable increase of the HSF systems fraction at $z<0.7$ (panel b). Irregular types harbour enhanced SFRs typically (both spheroidal and disk systems), coherently with their merger-related nature (panels c and d in the figure). The percentage of HSF objects in these types has not changed at $0.3<z<1.5$ within errors, the fraction of HSF objects in ID's being much higher than in IS's at all redshifts ($\sim 80$\% for ID's and $\sim 50$\% for IS's). This is normal if we consider that the former ones correspond to intermediate stages of gas-rich mergers, while the later ones group gas-poor major mergers and post-merger phases of the gas-rich ones. 

Figure\,\ref{Fig:mip_p_classification} implies that, at $0.7<z<1.5$, the enhanced star formation in red galaxies is mostly hosted by major mergers; mostly by gas-rich merger remnants at intermediate phases (ID's, which represents $\sim 30$\% of all red galaxies at these redshifts, see Fig.\,\ref{Fig:fractypes}) and by $\sim 50$\% of spheroidal remnants (IS's, which represent $\sim 30$\%).  At $z<0.7$, enhanced SFRs can be found still in most red major mergers (which represent $\sim 25$\% of the whole red population, see Fig.\,\ref{Fig:fractypes}) and in nearly all red RD's ($\sim 15$\% of it). RS's are the types that host the lowest levels of SFR enhancement at all redshifts, hosting $\sim 20$\% at most at $z<0.7$ (despite being $\sim 60$\% of the whole red population at $z<0.7$). 

Considering that our detectability timescale for gas-rich mergers is $\tau \sim 1$\,Gyr (\S\ref{Sec:RSEvolution}), the merger rate derived from our data for ID's at $0.3 < z < 0.7$ is $\sim 2.5\times 10^{-4}$ Gyr$^{-1}$ Mpc$^{-3}$ (considering their number density from Fig.\,\ref{Fig:mip_p_classification}), which is an estimate that is in good agreement with the one reported by 
\citet{2011AJ....141...87C} for gas-rich mergers using a different methodology (see their Figure\,6). This strongly supports the identification of our ID's with intermediate stages of gas-rich major mergers.

\begin{figure*}
\begin{center}
\includegraphics*[width=0.33\textwidth,angle=0]{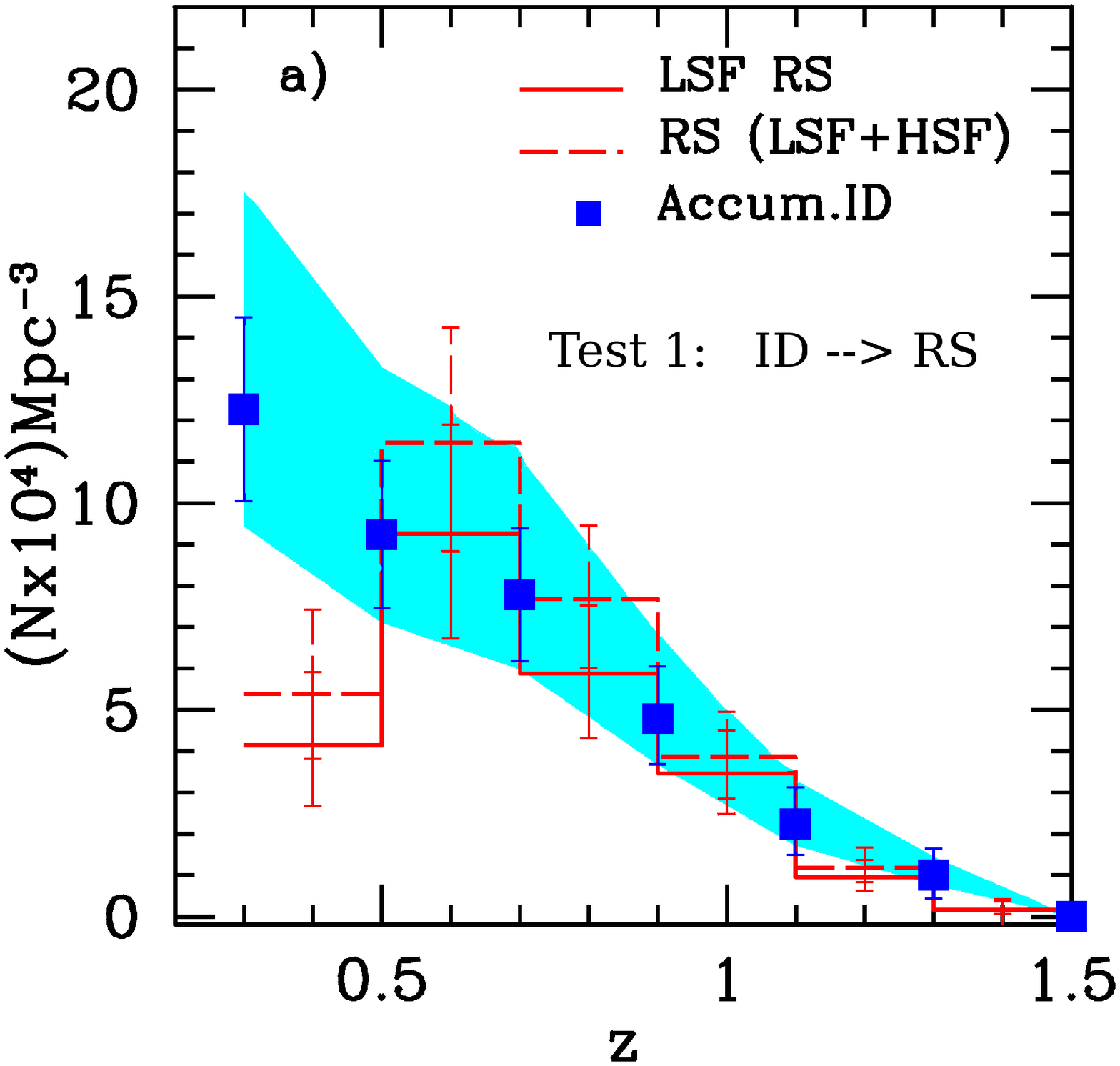}
\includegraphics*[width=0.33\textwidth,angle=0]{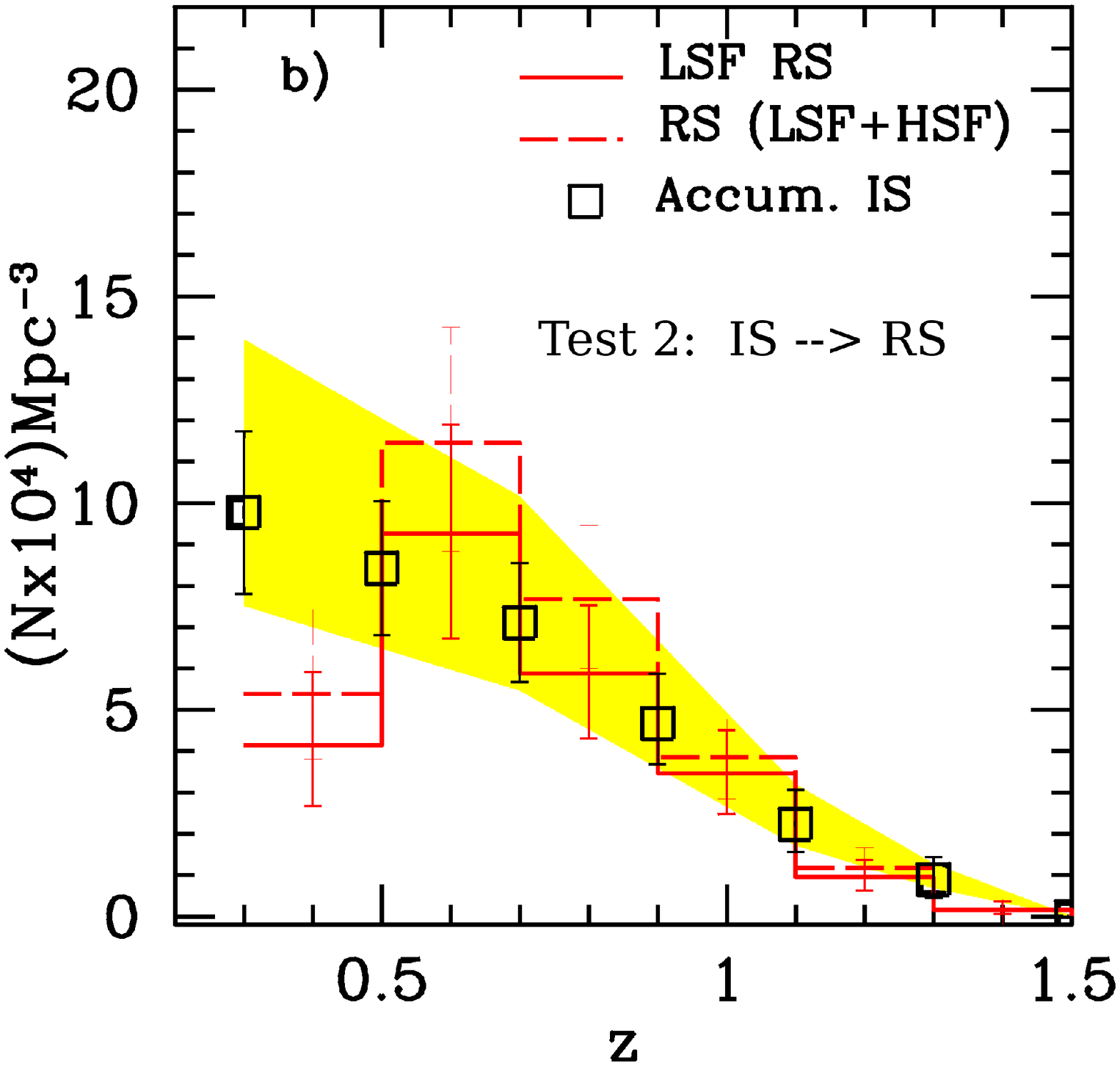}
\includegraphics*[width=0.33\textwidth,angle=0]{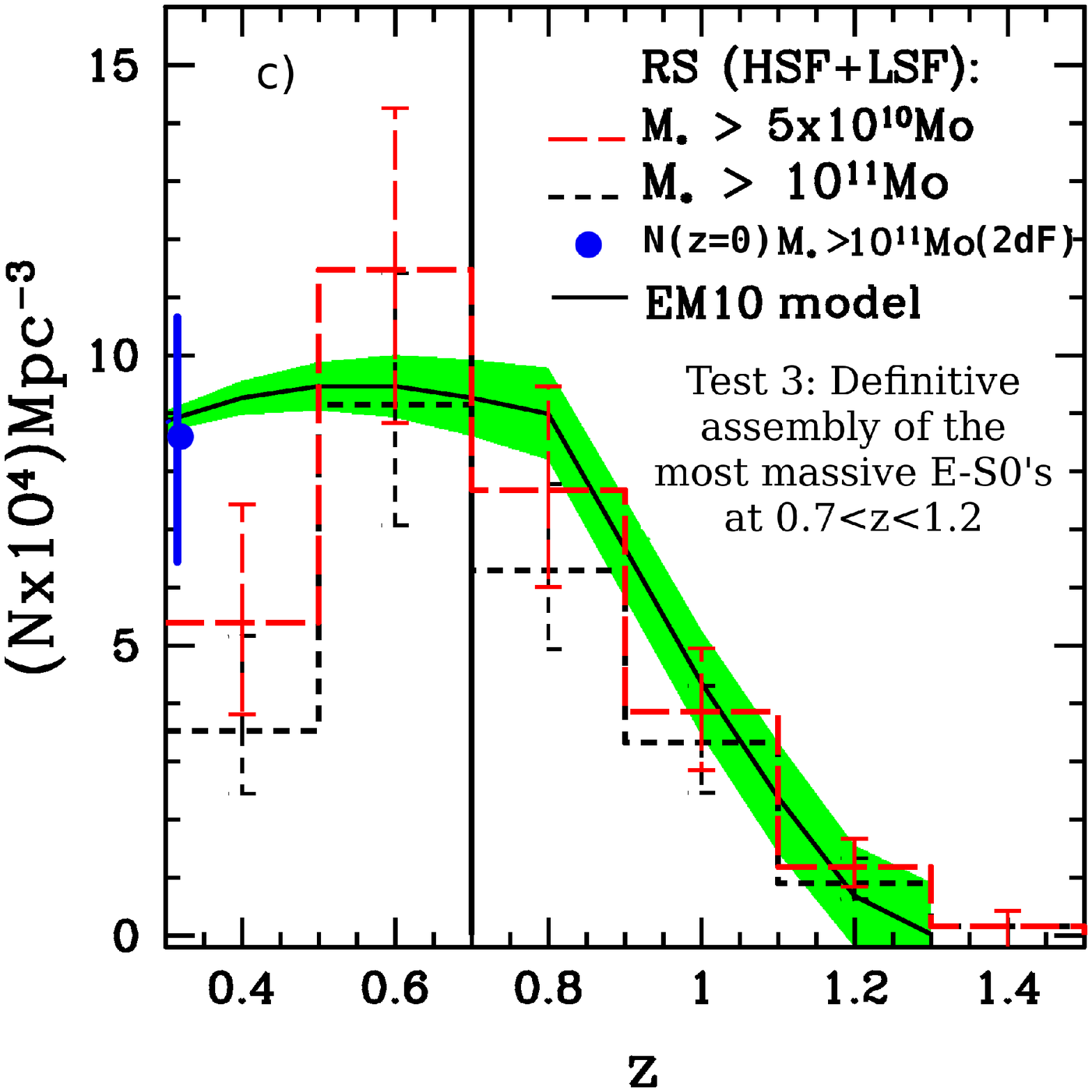}
\caption{Observational tests to the data of red galaxies for testing the hierarchical buildup of massive E-S0's, on the basis of the expectations of the EM10 model.  {\bf Panel a) Test 1\/}: Cumulative redshift distribution of red ID's (intermediate-to-late stages of gas-rich major mergers) since $z=1.5$ down to $z=0.3$, compared to the buildup of massive RS's (E-S0's) during the same time period. {\bf Panel b) Test 2\/}: Cumulative redshift distribution of red IS's (intermediate-to-late stages of gas-poor major mergers and late stages of gas-rich ones) since $z=1.5$ down to $z=0.3$, compared to the buildup of massive RS's (E-S0's) during the same time period. {\bf Panel c) Test 3\/}: Comparison of the redshift evolution of the number density of RS's (E-S0's) for two different mass ranges ($\mathrm{M}_*\gtrsim 5\times 10^{10}\Msun$ and $\mathrm{M}_*\gtrsim 10^{11}\Msun$) and the predictions of the EM10 model for E-S0's that end up with $\mathrm{M}_*\gtrsim 10^{11}\Msun$ at $z=0$. \emph{Vertical solid line}: Epoch at which the E-S0's with $\mathrm{M}_*\gtrsim 10^{11}\Msun$ at $z=0$ can be considered as definitively assembled according to the EM10 model. \emph{Blue circle}: Local number density of red galaxies with $\mathrm{M}_*\gtrsim 10^{11}\Msun$ derived by
\citet{2003ApJ...599..997M} from 2dF survey. 
The filled areas around the cumulative data points in panels a and b correspond to the uncertainties in the merger detectability timescales, while in panel c indicates the model uncertainties. The lowest redshift bin probably suffers from incompleteness.
}\label{Fig:acumIDSF}
\end{center}
\end{figure*}

\subsection{Tests to the definitive buildup of massive E-S0's through major mergers}
\label{Sec:TestsResults}

This section presents the results of the three observational tests commented in \S\ref{Sec:TestsResults} to test the hierarchical formation scenario of massive E-S0's.

\subsubsection{Test 1: Cumulative redshift distributions of red irregular disks since $z\sim 1.5$}
\label{Sec:Test1}

The first panel of Fig.\,\ref{Fig:acumIDSF} shows the result of the first test proposed in \S\ref{Sec:F07}. We plot the cumulative number density distributions of red ID's (intermediate stages of gas-rich major mergers, see Table\,\ref{Tab:correspondence}) since $z\sim 1.5$ down to redshift $z$ as obtained from eq.\,\ref{eq:density}, and compare it with the redshift evolution of the number density of red RS's (E-S0's, see the same table).  The detectability timescale considered for the ID's is that commented in \S\ref{Sec:RSEvolution}. The filled area around the cumulative data points accounts for the uncertainties in this timescale. The number densities at the lowest redshift bin are probably affected by volume and selection effects as suggested by this figure and as indicated in \S\ref{Sec:Comparison}, so we will override this redshift bin in all our results and comments henceforth.

The cumulative distribution of ID's reproduces pretty well both the redshift evolution of all RS's (E-S0's) and that of exclusively the LSF RS's (quiescent E-S0's) at $0.6<z<1.5$ within errors (most E-S0's at these redshifts are LSF's, see Fig.\,\ref{Fig:mip_p_classification}). This is compatible with that the bulk of red gas-rich major mergers at intermediate stages of their evolution at these redshifts evolve into the E-S0's that start to populate the Red Sequence at later cosmic times. Therefore, panel a in Fig.\,\ref{Fig:acumIDSF} supports the existence of the following evolutionary link among these two massive red galaxy types at $0.6<z<1.5$: 

\begin{equation}\label{eq:IDtoRS}
 \mathrm{\textbf{ID}} \rightarrow  \mathrm{\textbf{RS}}.  
\end{equation}

Panel a of Fig.\,\ref{Fig:acumIDSF} provides observational support to many aspects concerning the buildup of massive red galaxies expected by hierarchical models of galaxy formation:

\begin{enumerate}
 
\item At $0.6<z<1.5$ all irregular disks populating the massive end of the Red Sequence and nearby locations have evolved into the regular spheroids that appear on it some time later. 

\item This implies that the bulk of red massive E-S0's at $z\sim 0.6$ derive from disk galaxies that have migrated from the Blue Cloud to the Red Sequence during the time period enclosed at $0.6<z<1.2$.

\item It also supports that the dominant mechanism after this evolution has been major merging, as data is compatible with the fact that most massive E-S0's that have appeared on the Red Sequence since $z\sim 1.2$ seem to have undergone a previous phase of gas-rich major merger (i.e., a previous evolutionary stage as a red ID). 
\end{enumerate}

We remark that the major mergers detected at their phases as ID's by $z\sim 1.2$ in this study must have started $\sim 0.5$-0.7\,Gyr before, i.e., they were at their approaching (pre-merger) phases by $z\sim 1.5$. Therefore, the first generation of E-S0's that should result from these mergers according to the hierarchical scenario have appeared after $z\sim 1.2$. Accounting for this and for the fact that the EM10 model traces the evolution of massive E-S0's back-in-time up to $z=1.2$, a comparison between our data and this model is feasible only up to this redshift.

In conclusion, the panel a of Fig.\,\ref{Fig:acumIDSF} indicates that the data fulfill the first observational constraint imposed by the EM10 model (\S\ref{Sec:F07}), supporting the evolutionary track indicated in eq.~\ref{eq:track} (see also eqs.\ref{eq:track2} and \ref{eq:track3}). 

\subsubsection{Test 2: Cumulative redshift distributions of red irregular spheroids since $z\sim 1.5$}
\label{Sec:Test2}

Panel b of Fig.\,\ref{Fig:acumIDSF} now compares the accumulated redshift distribution of all IS's (gas-poor mergers plus post-merger stages of gas-rich ones, see Table\,\ref{Tab:correspondence}) with the redshift evolution of the number density of massive E-S0's (RS's) since $z\sim 1.5$. The detectability timescales considered for the IS's are those commented in \S\ref{Sec:RSEvolution}. The filled area around the cumulative data points accounts for the uncertainties in this timescale. Again, the accumulated distribution of IS's reproduces quite nearly the buildup of massive E-S0's at $0.6<z<1.5$ within errors. Therefore, this figure supports the existence of an additional evolutionary link at $0.6<z<1.5$ among the following red galaxy types: 

\begin{equation}\label{eq:IStoRS}
 \mathrm{\textbf{IS}} \rightarrow  \mathrm{\textbf{RS}}.  
\end{equation}

This panel of Fig.\,\ref{Fig:acumIDSF} also shows that data is compatible with the fact that most massive red E-S0's at $z\sim 0.6$ have also experienced a previous transitory phase as a red IS at $0.6<z<1.5$. The evolutionary links shown in eqs.\,\ref{eq:IDtoRS} and \ref{eq:IStoRS} can be summarized into the following evolutionary track:

\begin{equation}\label{eq:otro}
 \mathrm{\textbf{ID}} \rightarrow   \mathrm{\textbf{IS}} \rightarrow \mathrm{\textbf{RS}},
\end{equation}

\noindent which is the evolutionary path between red galaxy types expected by the hierarchical scenario of the EM10 model for the buildup of massive E-S0's (see eq.\,\ref{eq:track3}). Therefore, the panel b of Fig.\,\ref{Fig:acumIDSF} shows that the data also fulfill the second constraint imposed by the hierarchical evolutionary scenario proposed in the EM10 model (\S\ref{Sec:F07}), supporting the evolutionary track indicated in eq.~\ref{eq:track}.

\subsubsection{Test 3: Epoch of definitive assembly of massive E-S0's}
\label{Sec:Test3}

In the last panel of Fig.\,\ref{Fig:acumIDSF}, we compare the redshift evolution of the number density of E-S0's predicted by the EM10 model with the one obtained from our data for two different mass selections: for our nominal one ($\mathrm{M}_*>5\times 10^{10}\Msun$), and for an alternative one which is identical to our nominal one in all aspects, except in the mass range ($\mathrm{M}_*>10^{11}\Msun$).

The EM10 model traces back in time the evolution of the E-S0's that have $\mathrm{M}_*>10^{11}\Msun$ at $z=0$ (\S\ref{Sec:F07}). According to the model, these galaxies have been mostly assembled through major mergers at $0.7<z<1.2$, their buildup being frozen since then. Therefore, the model starts to trace the progenitors of these E-S0's at $z>0.7$, which have masses lower by a factor of $\sim 2$ compared to the E-S0's resulting from the merger. Consequently, the model predictions on the number density of E-S0's at $z>0.9$ can be compared with our results for $\mathrm{M}_*\gtrsim 5\times 10^{10}\Msun$ (as both studies trace similar mass ranges globally). But at lower redshifts, the EM10 model traces E-S0's that already have $\mathrm{M}_*>10^{11}\Msun$, so it is comparable to the results with a mass selection $\mathrm{M}_*>10^{11}\Msun$.  

As Fig.\,\ref{Fig:acumIDSF} c shows, at $z>0.7$, the model reproduces much better the settlement of the RS's with $\mathrm{M}_*\gtrsim 5\times 10^{10}\Msun$, whereas at $z<0.7$ it clearly follows the trend of RS's with $\mathrm{M}_*>10^{11}\Msun$, as expected from the arguments given above. However, our data seem to have completeness problems at $z<0.5$, so we cannot assure that the number density of E-S0's with $\mathrm{M}_*> 10^{11}\Msun$ has remained constant since $z\sim 0.7$. 

Nevertheless, the number density of E-S0's with $\mathrm{M}_*>10^{11}\Msun$ at $z\sim 0.6$ estimated with our data is
quite similar to the one estimated for these galaxies at $z=0$ \citep{2003ApJ...599..997M}, as shown in panel c of  Fig.\,\ref{Fig:acumIDSF}. This means that the number density of these objects has remained nearly constant since $z\sim 0.6$, in good agreement with the predictions of the EM10 model.

Although better data at $z<0.6$ are required to directly confirm this result, our data are coherent with the fact that E-S0's with $\mathrm{M}_*> 10^{11}\Msun$ have been definitively assembled since $z\sim 0.6$, supporting that the bulk of the assembly of red massive E-S0's has occurred during the $\sim 2.2$\,Gyr period elapsed at $0.7<z<1.2$, as predicted by the EM10 model. Therefore, we can conclude that our data fulfill the third constraint imposed by the hierarchical evolutionary scenario proposed in the EM10 model (\S\ref{Sec:F07}), supporting again the evolutionary track indicated in eq.~\ref{eq:track}. 

Summarizing, the three observational tests to the hierarchical evolutionary framework proposed by the EM10 model strongly supports this scenario, pointing to major mergers as the main mechanism for the buildup of massive E-S0's.

\section{Discusion}
\label{Sec:Discussion}

In general, our results strongly support a late definitive formation redshift for massive E-S0 ($z\lesssim1.5$), in agreement with hierarchical scenarios of galaxy formation (see references in \S\ref{Sec:introduction}). \citet{2010ApJ...709..644I} already claimed for the freezing in the assembly of massive E-S0's at $z\lesssim0.7$, proposing that a sudden drop of the gas-rich major merger rate must take place at $z\lesssim0.8$ to explain it. This fact was proven to be observationally feasible by the EM10 model, just accounting for the major merger fractions reported by observations \citep{2010arXiv1003.0686E}. The present study supports it observationally. 

Apart from the major mergers that seem to have been the main responsible mechanisms to place the massive E-S0's on the Red Sequence at $0.6<z<1.5$, other evolutionary processes must have contributed to the evolution of massive E-S0's down to the present (and even coeval with them). But, as commented above, they must not have risen their masses noticeably or changed their morphology noticeably, as E-S0's with $\log (M_*/\Msun)> 11$ seem to have been in place since $z\sim 0.6$ (see \S\ref{Sec:TestsResults}). This is supported by the fact that many of these processes are observed to be relevant only for the evolution of galaxies with lower masses \citep[in particular, the fading of stellar populations, see]{2002ApJ...577..651B,1998ApJ...496L..93D,2006A&A...458..101A}. 

We must also remark that our red galaxy sample does not trace the evolution of S0's observed in the clusters, which seems to have been relevant since $z\sim 0.4$-0.5 due to the environmental star-formation quenching of the spirals that fall into the clusters \citep{2007ApJ...660.1151D,2009ApJ...697L.137P,2009A&A...508.1141S,2011MNRAS.412..246V}. Cluster S0's have typical masses lower than those selected here \citep[$\mathrm{M}_*\lesssim 5\times 10^{10}\Msun$, see][]{1999ApJS..122...51D,2006MNRAS.373.1125B}, and hence, our results do not apply to them in general. Moreover, some studies indicate that the fraction of S0’s in groups is similar to that of clusters at $z < 0.5$ \citep{2009ApJ...692..298W}. Considering that most galaxies reside in groups \citep[$\sim 70$\%, see][]{2006ApJS..167....1B,2007ApJ...655..790C}, this means that the majority of S0’s in the Universe are located in groups (not in clusters). Nevertheless, note that this evolution in clusters does not contradict the EM10 model at all, because this model exclusively analyses the effects of the major mergers on galaxy evolution since $z\sim 1.2$, independently on the relevance of other evolutionary processes. 

To summarize, our study supports that major mergers have been the main drivers of the evolution of the massive end of the Red Sequence since $z\sim 1.5$, although other processes can also have contributed to it significantly at intermediate-to-low masses (especially, since $z\sim 0.6$). Our tests support observationally a late definitive buildup of the massive E-S0's through major mergers (mostly at $0.6<z<1.2$), in agreement with the expectations of hierarchical models of galaxy formation. 

\section{Conclusions}
\label{Sec:Conclusions} 

We study the buildup of the Red Sequence by analysing the structure, morphology, and star formation properties of a sample of red galaxies with stellar masses $\mathrm{M}_*\gtrsim 5\times 10^{10}\Msun$ at $0.3<z<1.5$. The novelty of this study is two-fold: first, our red galaxy sample includes galaxies both on the Red Sequence and at nearby locations on the Green Valley to trace transitory evolutionary stages towards it, and secondly, we have simultaneously considered structural (regular/irregular), morphological (disks/spheroids), and star formation enhancement (HSF/LSF) to define galaxy classes that can be directly associated with intermediate-to-late stages of major mergers, as well as with normal E-S0's and red regular disks. The redshift evolution of the fractions and number densities of each red galaxy type have been derived. Finally, these data are used to carry out a set of novel observational tests defined on the basis of the expectations of hierarchical models, to test the hierarchical origin of massive E-S0's. 

Both the number densities and fractions of regular galaxies increase with cosmic time at $0.3<z<1.5$, tracing the progressive buildup of massive normal early-type galaxies during the last $\sim 9$\,Gyr (mostly E-S0's, but also some red disks). The fractions of red regular spheroids (E-S0's) plus red disks increase by a factor of $\sim 6$ since $z\sim 1.3$ down to $z\sim 0.5$, whereas the fraction of irregulars (major mergers) decreases by the same factor. However, the number density of red irregulars (major mergers) is constant down to $z\sim 0.7$, decreasing at lower redshifts. This means that the fraction of red irregulars decreases only relatively to that of regulars, pointing to their transitory nature, as already claimed in previous studies. We find that the number density distributions of regular spheroids (E-S0's) and of irregulars (major mergers) cross at $z\sim 0.9$ for $\mathrm{M}_*\gtrsim 5\times 10^{10}\Msun$.

Enhanced SFRs compared to the average SFR of the global galaxy population at each redshift at $0.3 < z < 1.5$ are hosted by most irregular disks in our red sample and by nearly half of the irregular spheroids (in agreement with the major merger-related nature of both types). At $z\lesssim 0.7$, enhanced SF can also be found in nearly all red regular disks. On the other hand, only $\sim 25$\% of red regular spheroids (E-S0's) harbour enhanced star formation at all redshifts. 

The main result of this study is that we provide observational evidence pointing to the existence of a main evolutionary path among red galaxy types at $0.6<z<1.5$, being:\\[-0.1cm]

\noindent  {\bf Irregular disk $\rightarrow$ irregular spheroid $\rightarrow$  regular spheroid}. \\[-0.3cm]

This track traces the conversion of blue disks into passive E-S0's through major mergers and dominates the buildup of the Red Sequence at $z\gtrsim 0.6$, in excellent agreement with the expectations of the EM10 model. Our data are coherent with the prediction of this model about that the bulk of massive E-S0's with $\mathrm{M}_*\gtrsim 10^{11}\Msun$ have been definitively assembled through major mergers during the $\sim 2.2$\,Gyr period elapsed $0.6<z<1.2$, as also proposed by \citet{2010ApJ...709..644I}. 

Our results support observationally several expectations of hierarchical theories of galaxy formation concerning the buildup of massive red galaxies: 

\begin{enumerate}
 \item Data is compatible with the fact that the massive red regular galaxies at low redshifts derive from the irregular ones populating the Red Sequence and its neighbourhood at earlier epochs up to $z\sim 1.5$.

 \item The progenitors of the bulk of present-day massive red regular galaxies seem to have been disks that have migrated to the Red Sequence mostly through major mergers at $0.6<z<1.2$ (these mergers thus starting at $z\lesssim 1.5$).

 \item The formation of E-S0's that end up with $\mathrm{M}_* \gtrsim  10^{11}\Msun$ at $z=0$ through gas-rich major mergers seems to have frozen since $z\sim 0.6$.
\end{enumerate}

\section*{Acknowledgments}
The authors thank the anonymous referee for the provided input that helped to improve this publication significantly. We also thank M.~Bernardi, N.~Deveraux, S.~di Serego Alighieri, B.~Rothberg, and D.~Sobral for interesting and useful discussion on the topic. Supported by the Spanish Ministry of Science and Innovation (MICINN) under projects AYA2009-10368, AYA2006-12955, and AYA2009-11137, and by the Madrid Regional Government through the AstroMadrid Project (CAM S2009/ESP-1496, http://www.laeff.cab.inta-csic.es/\-projects/\-astromadrid/\-main/ index.php). Funded by the Spanish MICINN under the Consolider-Ingenio 2010 Program grant CSD2006-00070: "First Science with the GTC" (http://www.iac.es/\-consolider-ingenio-gtc/).  Based on observations made with the Isaac Newton and Willian Herschel Telescope operated on the island of La Palma by the Isaac Newton Group of Telescopes in the Spanish Observatorio del Roque de los Muchachos of the Instituto de Astrof\'{\i}sica de Canarias. Some of the data presented herein are part of the DEEP2 survey, funded by NSF grants AST95-09298, AST-0071048, AST-0071198, AST-0507428, and AST-0507483, as well as NASA LTSA grant NNG04GC89G. This work is based in part on services provided by the GAVO data center. S.D.H. \& G.

\bibliography{bib_v1.bib}
\bibliographystyle{mn2e}
\newpage
\parbox{0.45\textwidth}{
\noindent $^{1}$Instituto de Astrof\'{\i}sica de Canarias, C/ V\'{\i}a L\'actea, E-38200, La Laguna, Canary Islands, Spain\\
$^{2}$Departamento de Astrof\'{\i}sica, Universidad de La Laguna, Avda.\ Astrof\'{\i}sico Fco.\ S\'anchez, E-38200, La Laguna, Canary Islands, Spain\\
$^{3}$Departamento de Astrof\'{\i}sica, Facultad de CC. F\'{\i}sicas, Universidad Complutense de Madrid, E-28040 Madrid, Spain\\
$^{4}$Isaac Newton Group of Telescopes, Apartado 321, E-38700, Santa Cruz de La Palma, Canary Islands, Spain\\
$^{5}$Centro de Estudios de F\'{\i}sica del Cosmos de Arag\'{o}n, E-44001, Teruel, Spain\\
$^{6}$Instituto de Astrof\'{\i}sica de Andaluc\'{\i}a, IAA-CSIC, Apdo. 3044, E-18080 Granada, Spain\\
$^{7}$Max-Planck-Institut f\"{u}r Extraterrestrische Physik, Giessen-bachstrasse, D-85748, Garching, Germany\\
$^{8}$Universit\"{a}tssternwarte, Scheinerstrasse 1, D-81679, M\"{u}nchen, Germany\\
$^{9}$Department of Astronomy, 477 Bryant Space Center, University of Florida, Gainesville, FL, USA\\
$^{10}$Associate Astronomer at Steward Observatory, The University of Arizona, USA}

\label{lastpage}

\end{document}